\documentclass[pre,aps,twocolumn]{revtex4}
\usepackage{psfrag}
\usepackage{amssymb,amsmath,amsthm}
\usepackage[dvips]{graphicx}
\usepackage{verbatim}
\usepackage{epsfig}
\usepackage{epstopdf}
\usepackage{color}

\def\bnu{{\mbox{\boldmath $\nu $}}}

\newcommand{\DS}[1]{\textcolor{magenta} {\textsf{\small{[DS: #1]}}}}

\begin{document}
\title{
Magneto-optic dynamics in a ferromagnetic nematic liquid crystal}

\date{\today}

\author{
Tilen Potisk$^{1,2,*}$,
Alenka Mertelj$^3$,
Nerea Sebasti\'{a}n$^3$,
Natan Osterman$^{1,3}$, \\
Darja Lisjak$^3$,
Helmut R. Brand$^2$,
Harald Pleiner$^4$,
and Daniel Sven\v sek$^1$}
\affiliation{$^1$Department of Physics, Faculty of Mathematics and Physics, University of Ljubljana,  
SI-1000 Ljubljana,  Slovenia\\
$^2$Department of Physics, University of Bayreuth, 95440 Bayreuth, Germany \\
$^3$J. Stefan Institute, SI-1000 Ljubljana, Slovenia \\
$^4$Max Planck Institute for Polymer Research, 55021 Mainz, Germany\\
$^*$e-mail: tilen.potisk@uni-bayreuth.de }

\begin{abstract}
\noindent
We investigate dynamic magneto-optic effects in a ferromagnetic nematic
liquid crystal experimentally and theoretically.
Experimentally we measure the magnetization 
and the phase difference of the transmitted light when an external magnetic field is applied. 
As a model 
we study the coupled dynamics of the magnetization, $\bf M$,  and the director
field, $\bf n$, associated
with the liquid crystalline orientational order.
We demonstrate that the experimentally studied macroscopic dynamic
behavior reveals the importance of a dynamic cross-coupling between
$\bf M$ and $\bf n$. 
The experimental data are used to extract the value of the 
dissipative cross-coupling coefficient. 
We also make concrete predictions about how reversible cross-coupling terms between
the magnetization and the director could be detected experimentally by
measurements of the transmitted light intensity as well as by analyzing the
azimuthal angle of the magnetization and the director out of the plane
spanned by the anchoring axis and the external magnetic field.
We derive the eigenmodes of the coupled system and study their relaxation rates. 
We show that in the usual experimental set-up used for measuring the relaxation rates 
of the splay-bend or twist-bend eigenmodes of a nematic liquid crystal one 
expects for a ferromagnetic nematic liquid crystal a mixture of at least two eigenmodes.
\end{abstract}


\maketitle

\section{Introduction}

\noindent
In Ref.~\cite{bropgdg} Brochard and de Gennes suggested and discussed a ferromagnetic
nematic phase combining the long range nematic orientational order 
with long range ferromagnetic order in a fluid system. The synthesis and experimental 
characterization of ferronematics and ferrocholesterics, 
a combination of low molecular weight nematic liquid
crystals (NLCs) with magnetic liquids leading to a superparamagnetic phase,
started immediately \cite{rault} and continued 
thereafter \cite{amer,antonio87,antonio2004,kopcansky,ouskova,buluy,podoliak}. 
These studies made use of ferrofluids or magnetorheological fluids 
(colloidal suspensions of magnetic particles) \cite{rosensweig};
their experimental properties \cite{rosensweig,odenbach2004}
have been studied extensively in modeling 
\cite{mario2001,luecke2004,hess2006,sluckin2006,mario2008,klapp2015} 
using predominantly macroscopic descriptions 
\cite{mario2001,luecke2004,hess2006,mario2008}.

On the modeling side, the macroscopic dynamics of ferronematics 
was given first for a relaxed magnetization  \cite{jarkova2} 
followed by taking into account the magnetization as a dynamic degree of 
freedom \cite{jarkova} as well as incorporating chirality effects
leading to ferrocholesterics \cite{fink2015}. In parallel a Landau description 
including nematic as well as ferromagnetic order has been presented 
\cite{hpmhd}.
 
Truly ferromagnetic NLCs have been 
generated \cite{alenkanature} in 2013 followed by reports of further 
ferromagnetic NLCs in \cite{shuaiNC,smalyukhPNAS},
and their macroscopic static properties 
were characterized in detail \cite{alenkasoftmatter}. 
Quite recently ferromagnetic cholesteric liquid crystals have
been synthesized and investigated \cite{smalyukhPRL,smalyukhNatMat,alenkascience}.
For a review on ferromagnetic NLCs see Ref. \cite{alenkarev}.

In the present paper we describe in detail experimentally and theoretically 
the static and  dynamic properties of ferromagnetic NLCs \cite{tilenshort}.
We analyze the coupled dynamics of the magnetization 
and the director, initiated  and controlled by an external magnetic field. 
We show experimentally and theoretically that dissipative dynamic coupling terms
influence qualitatively the dynamics. Experimentally this is done 
by measuring the temporal evolution  
of the normalized phase difference associated with the dynamics of
the director.
Quantitative agreement between the experimental results 
and the model is reached and a dissipative cross-coupling coefficient 
between the magnetization and the director is accurately evaluated.   
It is demonstrated that this cross-coupling is crucial to account for
the experimental results thus underscoring the importance of such off-diagonal effects 
in this first multiferroic fluid system.
We also make concrete theoretical predictions of how the reversible dynamic cross-coupling
terms between magnetization and director influence the macroscopic dynamics and
how these effects can be detected experimentally.
The experimental and theoretical dynamic results 
discussed in some detail in this paper for 
low magnetic fields in ferromagnetic NLCs demonstrate the potential for
applications of these materials in displays and magneto-optic devices 
as well as in the
field of smart fluids.

The paper is organized as follows. In Section \ref{sec:experiments} 
we describe the experimental set-up
followed in Section \ref{sec:model} by the macroscopic model. The connection between the
measurements and the model is established in Section \ref{sec:connection}.
In Section \ref{sec:statics} we analyze the statics and in Sections \ref{sec:ON} 
and \ref{sec:OFF} we analyze in detail the coupled macroscopic dynamics of the 
magnetization and the director field when switching the external magnetic field on 
and off, respectively. Section \ref{sec:flucts} is dedicated to a theoretical 
analysis of fluctuations and
light scattering and in the conclusions we give a summary of the main results and a perspective.

\section{Experiments}
\label{sec:experiments}

\noindent
The experimental samples 
have been prepared along 
the lines described in detail in Refs.~\cite{alenkanature,alenkasoftmatter}.  
In brief, the BaSc$_x$Fe$_{12-x}$O$_{19}$ nanoplatelets were suspended in the liquid crystal mixture E7 (Merck, nematic -- isotropic transition temperature $T_{\mathrm{NI}}=58\,^\circ$C). The suspension was filled in liquid crystal cells with rubbed surfaces (thickness $d=20\,\mu$m, Instec Inc.), which induced homogeneous in-plane orientation of the NLC.
The volume concentration of the magnetic platelets in the nematic
low-molecular-weight liquid crystal E7 (Merck) has been estimated to be $\sim 1.3 \times 10^{-3}$ 
from the measurements of the magnetization magnitude \cite{alenkasoftmatter} which was $M_0 \sim 200$\,A/m. E7 suspensions show long term stability, with a homogeneous response to magnetic fields and no aggregates for a period of several months. 
A surfactant (dodecylbenzene sulfonic acid) was used for the treatment of the nanoplatelets which favors a perpendicular orientation of the NLC molecules with respect to the nanoplatelets. 
Quantitative values for the Frank coefficients for E7 are available in
the literature \cite{km}.



\begin{figure}[htb]
\includegraphics[width=3.5in]{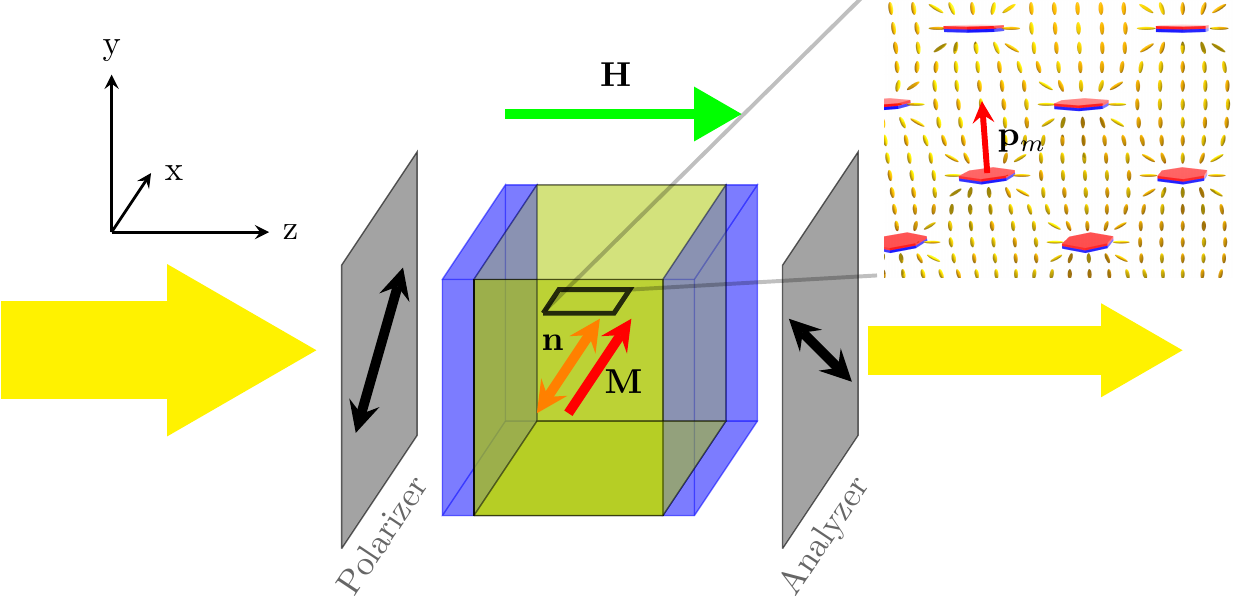}\vspace{2mm}
\includegraphics[width=3.35in]{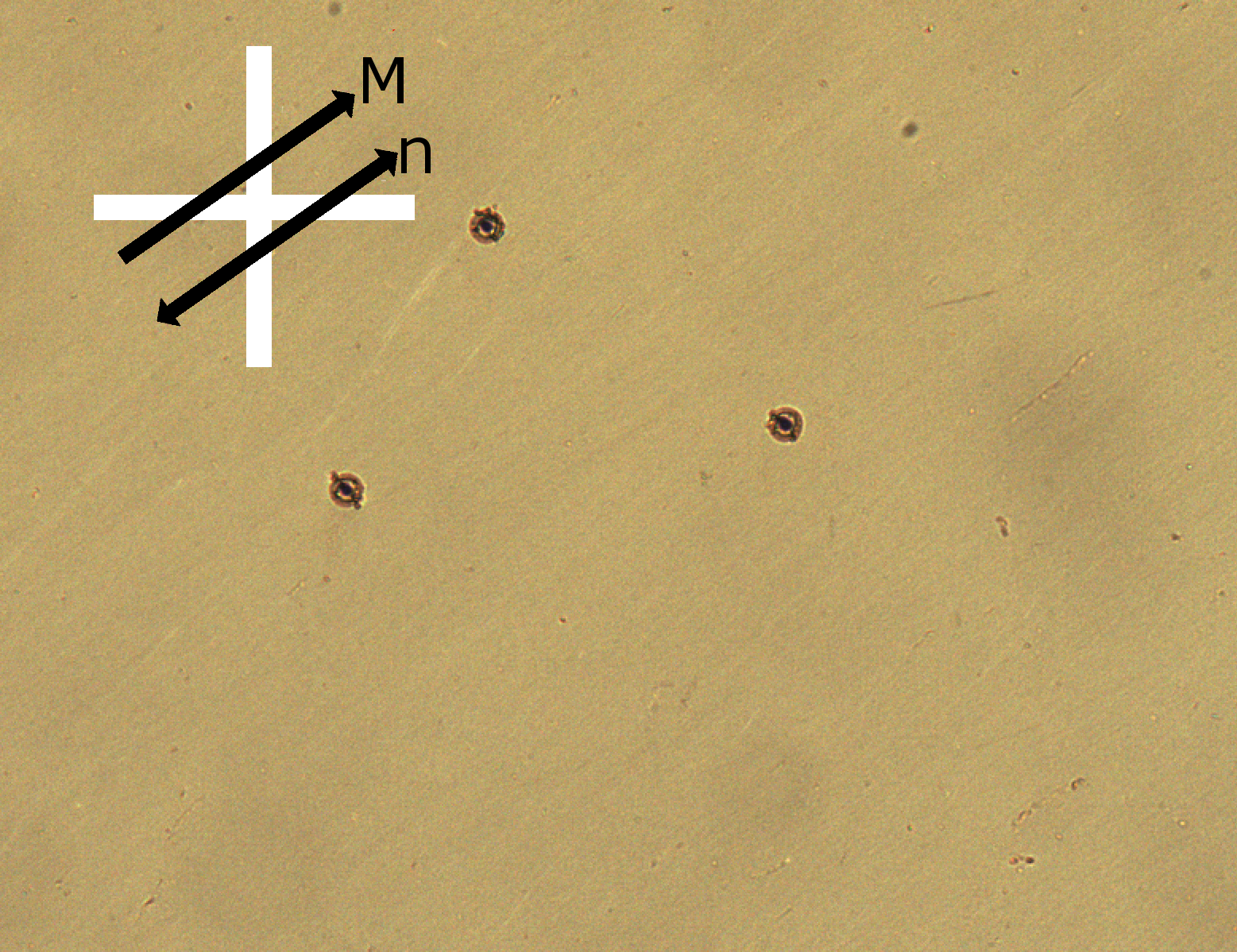}
\caption{ (Color online) Top: sketch of the experimental set-up and definition of
coordinate axes \cite{tilenshort}. The thick yellow arrows indicate the direction of
the light passing through the polarizer and the analyzer. In the absence of an applied magnetic field
(${\bf H}$, $z$ direction), the equilibrium director (${\bf n}$)
and magnetization (${\bf M}$) fields are only slightly pretilted from the $x$ direction.
Inset: distortion of the NLC director (ellipsoids, schematic) prevents flocculation of the suspended nanoplatelets carrying a magnetic moment ${\bf p}_m$ parallel to $\bf n$ in equilibrium.
Bottom: ferromagnetic E7 nematic 20\,$\mu$m sample placed between crossed polarizers, with the director at an angle of 45 degrees. Polarizing optical microscopy image width corresponds to 700\,$\mu$m. The spheres are cell spacers.
} 
\label{Bild0}
\end{figure}
Dynamics of the director was measured by inducing director reorientation in planarly treated 20\,$\mu$m cells (pretilt in the range 1-3 degrees) when applying a magnetic field perpendicular to the cell plates, Fig.~\ref{Bild0} (top). Experiments were performed on monodomain samples (see Ref.~\cite{alenkarev} for a description of monodomain sample preparation) so that the director is initially at 45 degrees with respect to the crossed polarizers, Fig.~\ref{Bild0} (bottom). Using polarizing microscopy, the monochromatic light intensity transmitted through the sample was recorded with a CMOS camera (IDS Imaging UI-3370CP, 997 fps) as a function of time on switching the magnetic field on and off. An interference filter (623.8\,nm) was used to filter the light from the halogen lamp used in the microscope. The transmitted light intensity is related to the phase difference between the ordinary and the extraordinary light as will be explained below. The advantage of using polarization microscopy is that the measurements are performed in the homogeneous region of the sample without spacers or other impurities. Recording the image of the sample during the measurements also allows us to simultaneously monitor the homogeneity of the response.


With the use of a vibrating sample magnetometer \cite{alenkasoftmatter} (LakeShore 7400 Series VSM) also the equilibrium $z$ component of the magnetic moment of the sample is measured. We note that this technique is not suitable for measuring the magnetization dynamically, as several seconds per measurement are required for ambient magnetic noise averaging.

\section{Macroscopic model}
\label{sec:model}

\noindent
Throughout the present paper 
we take into account the magnetization
${\bf M}$ and the director field $ {\bf n}$
as macroscopic variables;
in the following we focus
on the essential ingredients of their dynamics necessary to capture the experimental
results we will discuss. 
That is we assume isothermal conditions and discard flow effects.
For a complete set of macroscopic dynamic equations for ferronematics
we refer to Ref.~\cite{jarkova}.

The static behavior is described by the free energy density $f({\bf M}, {\bf n}, \nabla{\bf n})$,
\begin{equation}
       f = - \mu_0{\bf M}\cdot{\bf H} - {\textstyle{1\over 2}}A_1 ({\bf M}\cdot{\bf n})^2
              + {\textstyle{1\over 2}}A_2 \left(|{\bf M}|-M_0\right)^2 
              + f^F, \label{f}
\end{equation}
where $\mu_0$ is the magnetic constant, ${\bf H} = H \hat{\bf e}_z$ is the applied magnetic field, and $A_{1,2}>0$ will be assumed constant. The first term represents the coupling of 
the magnetization and the external magnetic field. 
Since $H\gg M_0$, the local magnetic field is equal to $\bf H$, which is fixed externally, and is thus independent of the ${\bf M}({\bf r})$ configuration.
The second term describes the static coupling between the director field and the magnetization (originating from the magnetic particles). The third term describes the energy connected with the deviation of the modulus of the magnetization from $M_0$. 
The last term is the Frank elastic energy associated with director distortions
\cite{degennesbook}
\begin{eqnarray}
        f^F &=& {\textstyle{1\over 2}}K_1(\nabla\cdot{\bf n})^2
                 + {\textstyle{1\over 2}}K_2\left[{\bf n}\cdot(\nabla\times{\bf n})\right]^2\nonumber \\
                 &+& {\textstyle{1\over 2}}K_3\left[{\bf n}\times(\nabla\times{\bf n})\right]^2, \label{f^F}
\end{eqnarray}
with positive elastic constants for splay ($K_1$), twist ($K_2$), and bend ($K_3$).
The saddle-splay elastic energy \cite{degennesbook} 
is zero in the considered geometry.
%
%
%
While it is a
good approximation to assume that $|{\bf M}|=M_0$,
we will take into account small 
variations of $|{\bf M}|$ (corresponding to large values of $A_2$).

The anchoring of the director at the plates is taken into account
using a finite surface anchoring energy \cite{rapini} 
\begin{equation}
        f^S = - {\textstyle{1\over 2}} W ({\bf n}_S\cdot{\bf n})^2,
        \label{f^S}
\end{equation}
where $W$ is the anchoring strength and ${\bf n}_{S}=\hat{{\bf e}}_{z}\sin\varphi_{s}+\hat{{\bf e}}_{x}\cos\varphi_{s}$
is the preferred direction specified by the director pretilt angle
$\varphi_{s}$. 

For the total free energy we have $F = \int\!\!f\,{\rm d}V+\int\!\!f^S\,{\rm d}S$ and the equilibrium condition requires $\delta F = 0$.

The macroscopic dynamic equations for the magnetization and the director read
\cite{pleinerbrandchapter,jarkova}
\begin{eqnarray}
        \dot M_i + X_i^R + X_i^D &=& 0, \label{Mdot}\\
        \dot n_i + Y_i^R + Y_i^D &=& 0, \label{ndot}
\end{eqnarray}
where the quasi-currents have been split into reversible ($X_i^R$, $Y_i^R$) 
and irreversible, dissipative ($X_i^D$, $Y_i^D$) parts. The reversible (dissipative) 
parts have the same (opposite) behavior under time reversal as the time derivatives of 
the corresponding variables, i.e, Eqs.~(\ref{Mdot})-(\ref{ndot}) are invariant under time 
reversal only if the dissipative quasi-currents vanish.

The quasi-currents are expressed as linear combinations of conjugate quantities 
(thermodynamic forces); they take the form
\begin{eqnarray}
        h_i^M &\equiv& {\delta f\over\delta M_i} = {\partial f\over\partial M_i}, \label{h^M}\\
        h_i^n &\equiv& \delta_{ik}^\perp {\delta f\over\delta n_k} = 
        \delta_{ik}^\perp\left({\partial f\over\partial n_k}-\partial_j\Phi_{kj}\right), \label{h^n}
\end{eqnarray}
with $\Phi_{kj} = {\partial f/\partial (\partial_j n_k)}$
and where the transverse Kronecker delta 
$\delta_{ik}^\perp = \delta_{ik} - n_i n_k$ projects onto the plane perpendicular 
to the director owing to the constraint ${\bf n}^2=1$.

In Ref.~\cite{tilenshort} we focused on the dissipative quasi-currents as they had a direct 
relevance for the explanation of the experimental results discussed there. 
In the present paper we also include the reversible 
quasi-currents, which give rise to transient excursions of $\bf M$ and $\bf n$ out of the switching plane. 

The dissipative quasi-currents take the form \cite{jarkova}
\begin{eqnarray}
\label{dyndis}
        X_i^D &=& b_{ij}^D h_j^M + \chi_{ji}^D h_j^n, \label{X}\\
        Y_i^D &=& {1\over\gamma_1} h_i^n + \chi_{ij}^D h_j^M, \label{Y}
\end{eqnarray}
with
\begin{eqnarray}
        \chi_{ij}^D &=& \chi_1^D\delta_{ik}^\perp M_k n_j + \chi_2^D\delta_{ij}^\perp M_k n_k,\label{chiD}\\
        b_{ij}^D &=& b_\parallel^D n_i n_j + b_\perp^D \delta_{ij}^\perp \label{bijD}
\end{eqnarray}
Throughout the present paper 
we will discard the biaxiality of the material which arises for  
${\bf n}\nparallel{\bf M}$.


The reversible quasi-currents are obtained by requiring that the entropy production $Y_ih_i^n+X_ih_i^M$ is zero \cite{jarkova}:
\begin{eqnarray}
        X_i^R &=& b_{ij}^R h_j^M + \chi^R \epsilon_{ijk} n_j h_k^n, \label{XR}\\
        Y_i^R &=& (\gamma_1^{-1})_{ij}^R h_j^n + \chi^R \epsilon_{ijk} n_j h_k^M, \label{YR}
\end{eqnarray}
where \cite{jarkova2}
\begin{eqnarray}
        b_{ij}^R &=& b_1^R \epsilon_{ijk} M_k + b_2^R \epsilon_{ijk} n_k n_p M_p \label{b_ij^R} \\
                           &&+ b_3^R(\epsilon_{ipq}M_p n_q n_j-\epsilon_{jpq}M_p n_q n_i),\nonumber\\
        (\gamma_1^{-1})_{ij}^R &=& 
	(\gamma_1^{-1})_1^R \epsilon_{ijk} n_k n_p M_p\label{gamma_inv} \\ 
	&& + 
	(\gamma_1^{-1})_2^R (\epsilon_{ijp} + \epsilon_{ipk} n_k n_j
	-  \epsilon_{jpk} n_k n_i ) M_p \nonumber.
\end{eqnarray}



For solving the system Eqs.~(\ref{Mdot})-(\ref{ndot}) a simple numerical method was used. We first discretized space into slices of width $\Delta z=d/(N-1)$, where $N$ is the number of discretization points. Empirically it was found that using $N=50$ is already sufficient. After discretizing space one obtains $N$ ordinary differential equations. 
Due to its simplicity, we use the Euler method. One step of the Euler method for the i-th component of the director field at $z$ is
\begin{equation}
n_i(t+\delta t,z)=n_i(t,z)-\delta tY_i(t,z)+\mathcal{O}(\delta t^2),
\end{equation}
where $\delta t$ is the time step. An analogous equation holds for the magnetization field and the equations are solved simultaneously. Since the numerical scheme for the director field is not norm preserving, we normalize the director field after each time step: $n_i \rightarrow n_i/\sqrt{n_jn_j}$.

In the discrete version, the two surface points are best treated by satisfying 
the same dynamic equations Eqs.~(\ref{Mdot})-(\ref{ndot}) as the internal points, with the 
addition of the surface anchoring energy Eq.~(\ref{f^S}) expressed as a volume density. 
The divergence part of the force Eq.~(\ref{h^n}) is then replaced by its surface flux 
(the volume density thereof, again):
\begin{equation}
        h_i^{n\, {\rm surf.}} = \delta_{ik}^\perp\left[{\partial f\over\partial n_k}+{1\over\Delta z}
                               \left(\nu_j\Phi_{kj}+{\partial f^S\over\partial n_k}\right)\right],
\end{equation}
where $\bnu$ is the surface normal pointing down (up) 
at the bottom (top) plate.

\section{Connection between measurements and the model}
\label{sec:connection}

\noindent
In equilibrium the magnetic-field-distorted director and magnetization fields are lying in the $xz$ plane, $\mathbf{n}=(\sin \theta, 0 , \cos \theta)$ and $\mathbf{M}=M(\sin \psi, 0, \cos \psi)$. 
In the absence of the magnetic field, the director is tilted from the $x$ axis by the pretilt $\varphi_s$, Eq.~(\ref{f^S}). The coordinate system used here is shown in Fig.~\ref{Bild0}. 
As explained earlier, the average $z$ component of the magnetization, $M_z$, is measured by the vibrating sample magnetometer. 
In modeling, it is obtained by averaging the $z$ component of the magnetization field,
\begin{equation}
\label{eq0}
M_z=\frac{1}{d}\int_{0}^d M\cos \psi(z)\, \mathrm{d}z.
\end{equation}

To derive the expression for the phase difference we start with an electric field $\mathbf{E}$, which is linearly polarized after the light passes through the polarizer,
\begin{equation}
\mathbf{E}=E_0{\mathbf{j}}\, {\rm e}^{\mathrm{i}(\mathbf{k}_i\cdot\mathbf{r}-\omega t)}, 
\end{equation}
where $E_0$ is the electric field amplitude, ${\mathbf{j}}$ the initial polarization, $\mathbf{k}_i$ the wave vector and $\omega$ the frequency of the incident light. In our case the wave vector points in the $z$ direction,$\mathbf{k}_i=k_0 \hat{\mathbf{e}}_z$, with $k_0=\frac{2\pi}{\lambda}$ being the wave number. The polarization of the light therefore lies in the $xy$ plane and is described by the two-component complex vector ${\bf j}=j_x(z)\hat{\bf e}_x+j_y(z)\hat{\bf e}_y$. 
As the light passes through the sample also the components of this (Jones) polarization vector change and we analyze these changes using the Jones matrix formalism (assuming perfectly polarized light) \cite{hechtbook}. 

The incident light first goes through the polarizer oriented at $45^\circ$ with respect to the $x$ axis, Fig.~\ref{Bild0}, and is linearly polarized with the initial Jones vector being $\mathbf{j}=\frac{1}{\sqrt{2}}(1,1)^T$.  The optical axis is parallel to the director and generally varies through the cell.
For any ray direction we can decompose the polarization into a polarization perpendicular to the optical axis (ordinary ray) and a polarization which is partly in the direction of the optical axis (extraordinary ray). The ordinary ray experiences an ordinary refractive index $n_o$ and the extraordinary ray experiences a refractive index $n_e$,
\begin{equation}
	\label{refract}
	n_e^{-2}(z)=n_{e0}^{-2}\sin^2\theta(z)+n_o^{-2}\cos^2\theta(z),
\end{equation}
where $n_{e0}$ is the extraordinary refractive index.

To calculate the intensity of the transmitted light, one first divides the liquid crystal cell into $N$ 
thin slices of width $h=d/N$ and describes the effect of each slice on the polarization by the phase matrix
\begin{equation}
\textsf{W}(z) = 
 \begin{pmatrix}
 {\rm e}^{\mathrm{i}k_0[n_e(z)-n_o]h/2} &0 \\
0 &  {\rm e}^{-\mathrm{i}k_0[n_e(z)-n_o]h/2}
 \end{pmatrix}.
\end{equation}
In the limit $N\to \infty$ we can express the transmission matrix of the liquid crystal cell as
\begin{equation}
\label{transm}
\textsf{T} = 
 \begin{pmatrix}
 {\rm e}^{\mathrm{i}\phi/2} &0 \\
0 &  {\rm e}^{-\mathrm{i}\phi/2}
 \end{pmatrix},
\end{equation}
where we have introduced the phase difference
\begin{equation}
	\label{eqfaza}
	\phi=k_0 \int_0^d [n_e(z)-n_o]\mathrm{d}z.
\end{equation}

In general, as we will see, the director can have also a nonzero component in the $y$ direction.
In this case the simple expression for the transmission matrix Eq.~(\ref{transm}) does not hold anymore and must be generalized.

We start the derivation of the general transmission matrix by assuming a general orientation of the director,
\begin{equation}
\mathbf{n}=(\sin\theta \cos \varphi, \sin\theta \sin \varphi, \cos\theta).
\end{equation}
The azimuthal angle of the director $\varphi$ can vary through the cell and the transformation matrix at point $z$ is
\begin{equation}
\textsf{T}(z) = 
\textsf{R}[-\varphi(z)]\textsf{W}(z)\textsf{R}[\varphi(z)],
\end{equation}
where $\textsf{R}$ is the rotation matrix 
\begin{equation}
\textsf{R}(\varphi) = 
 \begin{pmatrix}
 \cos(\varphi) & \sin(\varphi) \\
 -\sin(\varphi) &   \cos(\varphi)
 \end{pmatrix}.
\end{equation}
Our goal is to find the transfer matrix for the whole cell,
\begin{equation}
\label{eeq1}
\textsf{T}=\prod_{z \in [0,d]}^{\longleftarrow}\textsf{T}(z),
\end{equation}
where the arrow denotes the ordered product starting from $\textsf{T}(0)$ at the right side.
We first notice that 
\begin{equation}
\textsf{T}(z)\approx I+\mathrm{i} \frac{k_0[n_e(z)-n_o]h}{2} \begin{pmatrix}
 \cos[2\varphi(z)]& \sin[2\varphi(z)] \\
 \sin[2\varphi(z)] &   -\cos[2\varphi(z)]
 \end{pmatrix},
\end{equation}
where $I$ is the identity matrix.
Consequently we can write $\textsf{T}(z)$ as an exponential,
\begin{equation}
\textsf{T}(z) = \lim_{h\to 0}\exp [\mathrm{i}\textsf{A}(z)h],
\end{equation}
where $\textsf{A}$ is defined by
\begin{equation}
\textsf{A}(z)=\frac{k_0[n_e(z)-n_o]}{2} \begin{pmatrix}
 \cos[2\varphi(z)] & \sin[2\varphi(z)] \\
 \sin[2\varphi(z)] &   -\cos[2\varphi(z)]
 \end{pmatrix}.
\end{equation}
We can now rewrite Eq.~(\ref{eeq1}) as
\begin{equation}
\label{eeq2}
\textsf{T}= \lim_{h\to 0}\exp \left[\mathrm{i}\sum_{z\in [0,d]}\textsf{A}(z)h\right] = \exp [\mathrm{i}\int_{0}^d \textsf{A}(z)\mathrm{d}z],
\end{equation}
where we used
\begin{equation}
{\rm e}^{\textsf{A}h}{\rm e}^{\textsf{B}h}={\rm e}^{(\textsf{A}+\textsf{B})h}+\frac{1}{2}[\textsf{A},\textsf{B}]h^2 + \mathcal{O}(h^3).
\end{equation}
The exponential of the $2\times2$ matrix from Eq.~(\ref{eeq2}) reads
\begin{equation}
\textsf{T}= \begin{pmatrix}
 \cos(c)+\mathrm{i}\frac{a}{c}\sin(c) &\mathrm{i}\frac{b}{c}\sin(c) \\
\mathrm{i}\frac{b}{c}\sin(c) &   \cos(c)-\mathrm{i}\frac{a}{c}\sin(c)
 \end{pmatrix},
\end{equation}
where $c=\sqrt{a^2+b^2}$ with 
\begin{equation}
\label{ab}
\begin{split}
a&=\frac{k_0}{2}\int_{0}^d [n_e(z)-n_o]\cos[2\varphi(z)]\mathrm{d} z, \\
b&=\frac{k_0}{2}\int_{0}^d [n_e(z)-n_o]\sin[2\varphi(z)]\mathrm{d} z.
\end{split}
\end{equation}
We then let the light pass through an analyzer $\textsf{P}_\alpha$ at an angle $\alpha$,
\begin{equation}
\textsf{P}_\alpha=\begin{pmatrix}
 \cos^2\alpha & \sin \alpha \cos \alpha \\
\sin \alpha \cos \alpha &   \sin^2\alpha
 \end{pmatrix},
\end{equation}
which gives for the final Jones vector ($\alpha=-45^\circ$)
\begin{equation}
	\mathbf{j}'=\frac{\mathrm{i}a\sin (c)}{\sqrt{2}c}\begin{pmatrix}
	1 \\
	-1 
 	\end{pmatrix}.
\end{equation}
This yields the measured normalized intensity
\begin{equation}
\label{normint}
\frac{I}{I_0}=\mathbf{j'^*}^T\mathbf{j'}=\frac{a^2}{c^2}\sin^2(c).
\end{equation}

Next we evaluate the relation between the phase difference and the measured intensity.
Let $\mathbf{j}$ be the Jones vector after the liquid crystal cell,
\begin{equation}
\mathbf{j}=\begin{pmatrix}
z_1{\rm e}^{\mathrm{i}\phi} \\
z_2
 \end{pmatrix},
\end{equation}
where $z_1$ and $z_2$ are real and $z_1^2+z_2^2=1$. Generally $|z_1|\neq |z_2|$. After an analyzer with $\alpha = -45^\circ$ we have a Jones vector
\begin{equation}
\mathbf{j}'=\frac{1}{2}(z_1{\rm e}^{\mathrm{i}\phi}-z_2)\begin{pmatrix}
1 \\
-1
 \end{pmatrix}
\end{equation}
and the intensity is related to the phase difference as
\begin{equation}
\frac{I}{I_0}=\frac{1}{2}\left[1-2z_1z_2\cos(\phi)\right].
\end{equation}
Only if the director is restricted to the $xz$ plane, 
$z_1=z_2$ and we have 
\begin{equation}
\frac{I}{I_0}=\frac{1}{2}\left[1-\cos(\phi)\right]=\sin^2\left(\frac{\phi}{2}\right),
\end{equation}
such that the relation between the intensity and the phase difference is
\begin{equation}
\label{eqff}
\phi=m\pi \pm 2\arcsin\left[\sqrt{\frac{I}{I_0}}\right],
\end{equation}
where $m\in \mathbb{Z}$ and the sign $\pm$ is determined by demanding that $\phi$ is sufficiently smooth. Generally however, the quantity obtained from the measured intensity by Eq.~(\ref{eqff}) is not the phase difference. It is the phase difference only when the director field is in the $xz$ plane. For the analysis of the dynamics not confined to the $xz$ plane, Sec.~\ref{sec:reversible}, we will therefore use the normalized intensity Eq.~(\ref{normint}).

In the case when the dynamics is in the $xz$ plane, 
to compare the numerical results with the experiments and also to compare the
dynamics of the director with the dynamics of the magnetization,
it is convenient to introduce the normalized phase difference
\begin{equation}
	r(H) = 1 - \frac{\phi(H)}{\phi_0},
\end{equation} 
where $\phi_0$ is the phase difference at zero magnetic field. The normalized phase difference is zero at $t=0$ and is always smaller or equal to 1. 
It can also assume negative values as we will see.

\section{Statics}
\label{sec:statics}

\noindent 
In this Section we present experimental and numerical results of statics and derive analytic formulae for the equilibrium configurations in the low and large external magnetic field limits.

\begin{figure}[htb]
\includegraphics[width=3.3in]{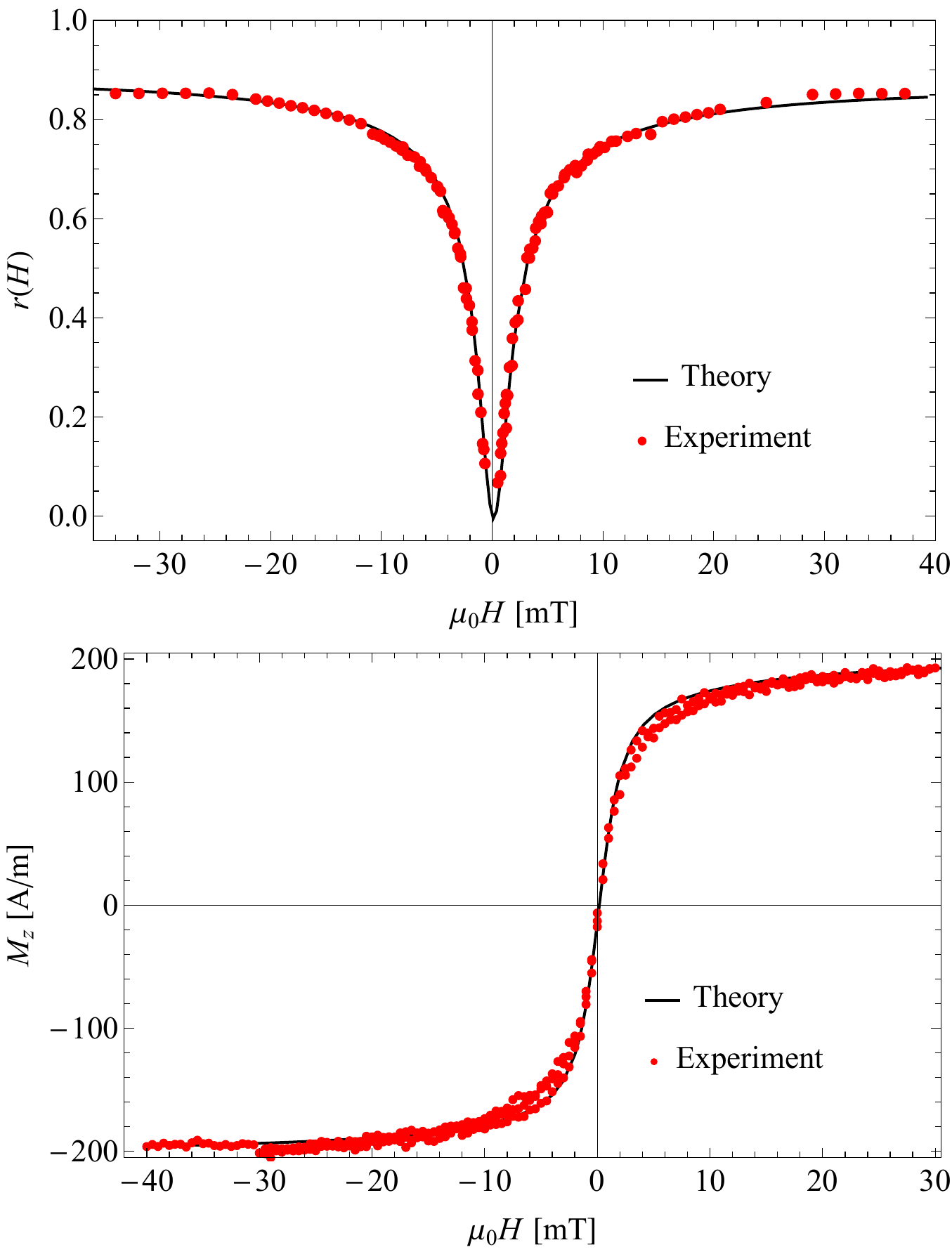}
\caption{ (Color online) Comparison of experimental and theoretical static results:
(top) normalized phase difference $r(H)$ and 
(bottom) magnetization component $M_z$ as functions of the
magnetic field $\mu_0 H$. 
} \label{Bild1}
\end{figure}

In Fig.~\ref{Bild1} we compare the numerical results of the equilibrium normalized phase difference to the experimental data. Below we will show in Eqs.~(\ref{up1})-(\ref{up11}) that the equilibrium normalized phase difference is quadratically dependent on the applied magnetic field at small magnetic fields.
The normalized phase difference saturates quickly above $\mu_0H$ = 10 mT at a value which is less than 1, which means there is a limit to how much the director field deforms. 
We also observe that the dependence of the equilibrium normalized phase difference is not symmetric with respect to the $\mu_0H=0$ axis, which is seen in experiments as well. The reason for this is the nonzero pretilt at both glass plates.

>From the fits to the model we extract values
for the anchoring strength $W$, 
the pretilt angle $\varphi_s$, the Frank elastic constant $K\equiv K_1=K_3$ in the one constant approximation, and the static coupling coefficient $A_1$:
\begin{eqnarray}
		W &\sim& 4 \times 10^{-5}\,{\rm J/m}^2, \label{Wextracted}\\
		\varphi_s &\sim& -0.05,\\
		K &\sim& 17\,{\rm pN},\\
		A_1 &\sim& 140 \mu_0.\label{A1extracted}
\end{eqnarray}
The extracted parameters Eqs.~(\ref{Wextracted})-(\ref{A1extracted}) correspond to the (local) minimum of the sum of squares of residuals between the numerical and experimental values of the normalized phase difference. This minimum was sought in sensible parameter ranges (for example the Frank elastic constant was sought in the range between 5\,pN and 25\,pN). There are several indications that this minimum is at least very close to the global one. First, the extracted value of the Frank elastic constant is close to the value of $K_3$ in the pure E7 NLC. Secondly, the extracted pretilt is within the range specified by the cell provider. Moreover the value of the static coupling is similar to that estimated for the ferromagnetic NLC based on 5CB \cite{alenkasoftmatter}.

The limiting behaviors of the normalized phase difference and the normalized $z$ component of the magnetization as the magnetic field goes to zero or infinity can be calculated analytically. 
In all cases the boundary condition is
\begin{equation}
K \frac{\partial \theta}{\partial z}\nu_z+\frac{\partial f^S}{\partial \theta}=0,
\end{equation}
where $\nu_z$ is the $z$ component of the surface normal pointing upwards at $z=d$ and downwards at $z=0$.

\subsection{Low magnetic fields}

\noindent
The free energy density in lowest order in deviations of magnetization and director field from the equilibrium is
\begin{equation} 
\label{free1}
f =\frac{1}{2}K \left(\frac{\partial \theta}{\partial z}\right)^2+\frac{1}{2}A_1M_0^2(\theta-\psi)^2+\mu_0HM_0\psi.
\end{equation} 
The equilibrium solutions for the angles are
\begin{align} 
\label{angles}
&\theta(z)=\frac{1}{2}\frac{\mu_0HM_0}{K}z(z-d) -\frac{\mu_0HM_0d}{2W}+ \frac{\pi}{2}-\varphi_s,\\
&\psi(z)=\theta(z)-\frac{\mu_0HM_0}{A_1M_0^2}.
\end{align} 
After inserting the solutions Eqs.~(\ref{angles}) in equations for the normalized phase difference and magnetization, one gets
%
%
\begin{align} 
\label{up1}
&r(H)=r_0 \frac{\mu_0HM_0d^2}{6K} \nonumber \\
&\times\left[\frac{\mu_0HM_0d^2}{20K}\left(1+10 \frac{\xi}{d}+30 \frac{\xi^2}{d^2}\right)+\left(1+6 \frac{\xi}{d}\right)\varphi_s\right],
\end{align} 
where $\xi=K/W$ is the so called anchoring extrapolation length and $r_0=n_{e0}(n_{e0}+n_o)/(2n_o^2)$.
In the limit of infinite anchoring the normalized phase difference reads
\begin{equation} 
	r(H)=r_0 \frac{\mu_0HM_0d^2}{6K}\left(\frac{\mu_0HM_0d^2}{20K}+\varphi_s \right).
	\label{up11}
\end{equation} 
One can also observe that the location of the minimum of the normalized phase difference is shifted to a value $\mu_0H_{\mathrm{min}}$ determined by the pretilt:
\begin{equation} 
\label{up2}
-\frac{10K\varphi_s\left(1+6\frac{\xi}{d}\right)}{M_0d^2 \left(1+10 \frac{\xi}{d}+30\frac{\xi^2}{d^2}\right)}\xrightarrow{W\to \infty}- \frac{10K\varphi_s}{M_0d^2}.
\end{equation} 
Eqs.~(\ref{up1}) and (\ref{up2}) are useful for determining the anchoring strength $W$ and the pretilt $\varphi_s$. 

>From the behavior of the normalized phase difference at low fields Eqs.~(\ref{up1})-(\ref{up11}) one cannot determine the value of the static coupling $A_1$. It can on the other hand be determined from the low-field behavior of the magnetization. 
In Fig.~\ref{Bild1} we see that the behavior is linear for low magnetic fields as can be shown analytically:
\begin{equation} 
\label{maglow}
\frac{M_z}{M_0}=\varphi_s+\left(\frac{1}{A_1M_0^2}+\frac{1}{12}\frac{d^2}{K}+\frac{d}{2W}\right)\mu_0H M_0.
\end{equation}

\subsection{Large magnetic fields}

\noindent

\noindent
In the large magnetic field limit we assume that both the polar angle of the director and the magnetization are either close to 0 if the applied magnetic field is positive ($+$) or close to $\pi$ if the applied magnetic field is negative ($-$).
The corresponding solutions will be denoted as $\theta^+(z),\theta^-(z),\psi^+(z)$, $\psi^-(z)$, $M_z^+$, $M_z^-$, $r^+$, and $r^-$.

The free energy in the case of a positive magnetic field is
\begin{equation} 
\label{free2}
f \approx\frac{1}{2}K \left(\frac{\partial \theta}{\partial z}\right)^2+\frac{1}{2}A_1M_0^2(\theta-\psi)^2+\frac{1}{2}\mu_0HM_0 \psi^2.
\end{equation} 
The equilibrium solutions for the angles $\theta^+(z)$ and $\psi^+(z)$ are
\begin{align} 
\label{angles2}
&\theta^+(z)=\frac{\frac{\pi}{2}-\varphi_s}{1+q\xi\tanh\left(\frac{qd}{2}\right)}\frac{\cosh\left[q(z-\frac{d}{2})\right]}{\cosh\left(\frac{qd}{2}\right)},\\
&\psi^+(z)=\frac{\theta^+(z)}{1+\frac{\mu_0|H|M_0}{A_1M_0^2}},
\end{align} 
where
\begin{equation}
	\label{qnum}
	q^2=q_0^2\frac{\mu_0|H|M_0}{\mu_0|H|M_0+A_1M_0^2}
\end{equation} 
with $q_0=\sqrt{A_1M_0^2/K}$ (which is proportional to the inverse ``magnetization coherence length'' of the director). 

The normalized $z$ component of the magnetization for large fields is
\begin{eqnarray}
\label{maghigh}
&&\frac{M_z^+}{M_0}=1-\\
&&\frac{[\frac{\pi}{2}-\varphi_s]^2(qd+\sinh(qd))A_1^2M_0^4}{4qd\left[1+q\xi\tanh\left(\frac{qd}{2}\right)\right]^2\cosh^2\left(\frac{qd}{2}\right)(A_1M_0^2+\mu_0|H|M_0)^2}\nonumber
\end{eqnarray}
and the normalized phase difference is
\begin{eqnarray}
\label{phasehigh}
&&r^+(H) = 1-\frac{n_or_\infty k_0d}{2\phi_0}\frac{\left[\frac{\pi}{2}-\varphi_s \right]^2}{\left[1+q\xi\tanh(\frac{qd}{2})\right]^2}\frac{qd+\sinh(qd)}{2qd\cosh(\frac{qd}{2})^2}\nonumber\\
&&-\frac{n_or_\infty k_0d}{4\phi_0}(3r_\infty/4-1/3)\frac{\left[\frac{\pi}{2}-\varphi_s \right]^4}{\left[1+q\xi\tanh(\frac{qd}{2})\right]^4}\nonumber\\ &&\times \frac{6qd+8\sinh(qd)+\sinh(2qd)}{8qd\cosh\left(\frac{qd}{2}\right)^4},
\end{eqnarray}
where $r_\infty=(n_{e0}^2-n_o^2)/n_{e0}^2$. 

It follows from symmetry that $\theta^-(\varphi_s) = \pi-\theta^+(-\varphi_s)$, $\psi^-(\varphi_s) = \pi-\psi^+(-\varphi_s)$, $M_z^-(\varphi_s) = -M_z^+(-\varphi_s)$, and $r^-(\varphi_s) = r^+(-\varphi_s)$

Since the magnetization is not anchored at the boundary, in Eq.~(\ref{maghigh}) it was sufficient to consider terms not higher than $(\psi^{+})^2$.
On the other hand, due to the anchoring of the director field, in Eq.~(\ref{phasehigh}) we expanded the phase difference to the order $(\theta^+)^4$. It should be noted, that the approximation for the phase difference is better if the anchoring $W$ is low, i.e., $q\xi \gg 1$ or $W \ll \sqrt{A_1M_0^2K}$. 


In the large magnetic field limit, where $qd \gg 1$, and if $q\xi \gg 1$ in addition, one can study asymptotic behavior of Eqs.~(\ref{maghigh}) and (\ref{phasehigh}):
\begin{eqnarray}
	r^+(H) &\asymp& r^+(\infty)-\frac{f^+(q_0)}{\mu_0|H|},\label{asympph}\\
	\frac{M_z^+}{M_0} &\asymp& 1-\frac{h^+(q_0)}{(\mu_0H)^2},\label{asympm}
\end{eqnarray}
where $f^+$ and $h^+$ are functions of static parameters for positive magnetic fields and $r^+(\infty) = \lim_{\mu_0H\to \infty}r^+(H)$.
The behavior of the magnetization $M_z$, Fig.~\ref{Bild1}, may at a first glance look like the Langevin function, often observed in magnetic systems. Eq.~(\ref{asympm}) tells us that this is not the case, since the Langevin function saturates with the first power in magnetic field, whereas here the saturation Eq.~(\ref{asympm}) is of second order in $H$. 

\begin{figure}[htb]
\includegraphics[width=3.3in]{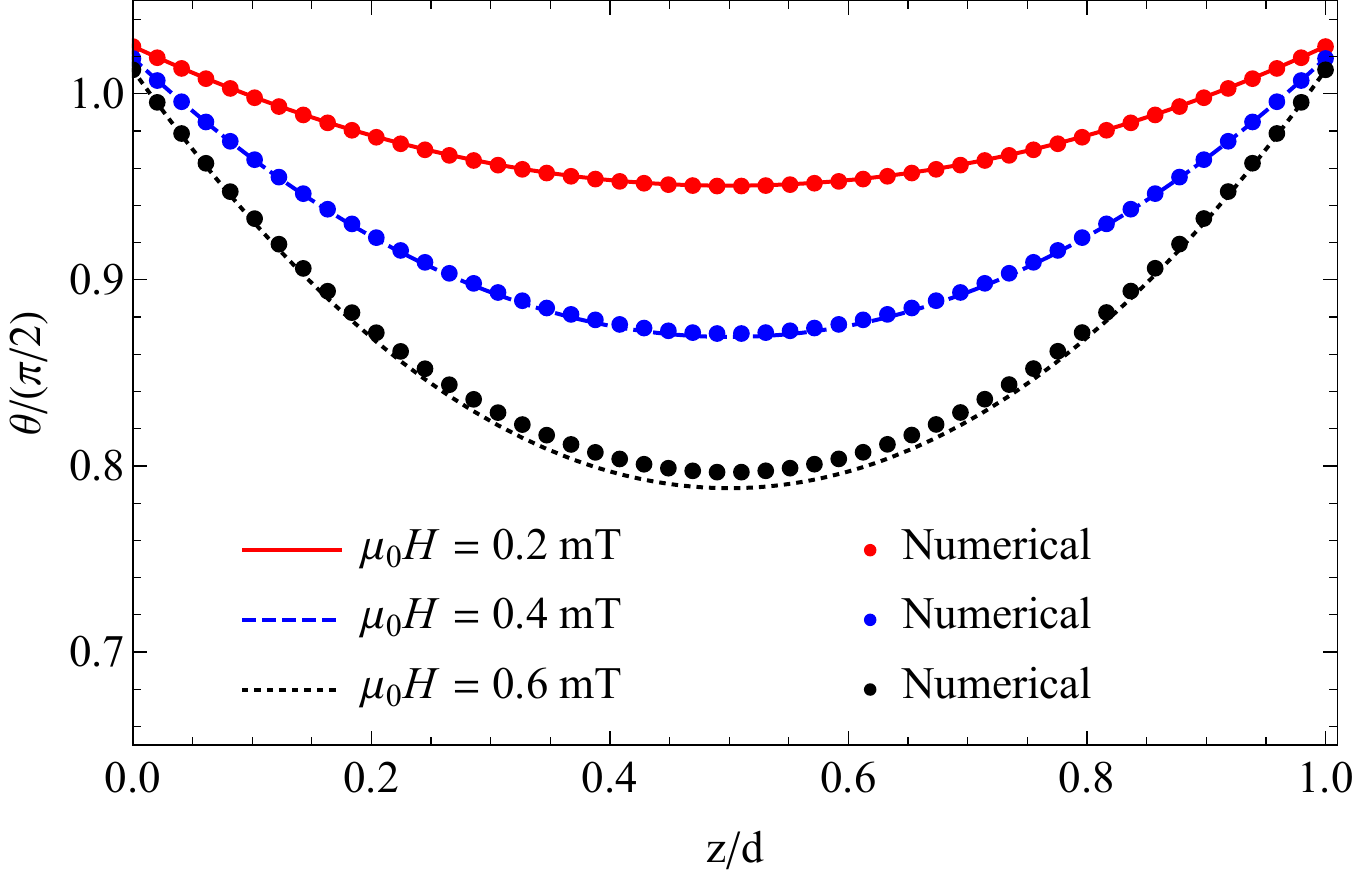}
\caption{(Color online) For low magnetic fields, the numerically calculated polar angle of the director is in agreement with Eq.~(\ref{angles}).}
\label{Bild3}
\end{figure}
\begin{figure}[htb]
\includegraphics[width=3.3in]{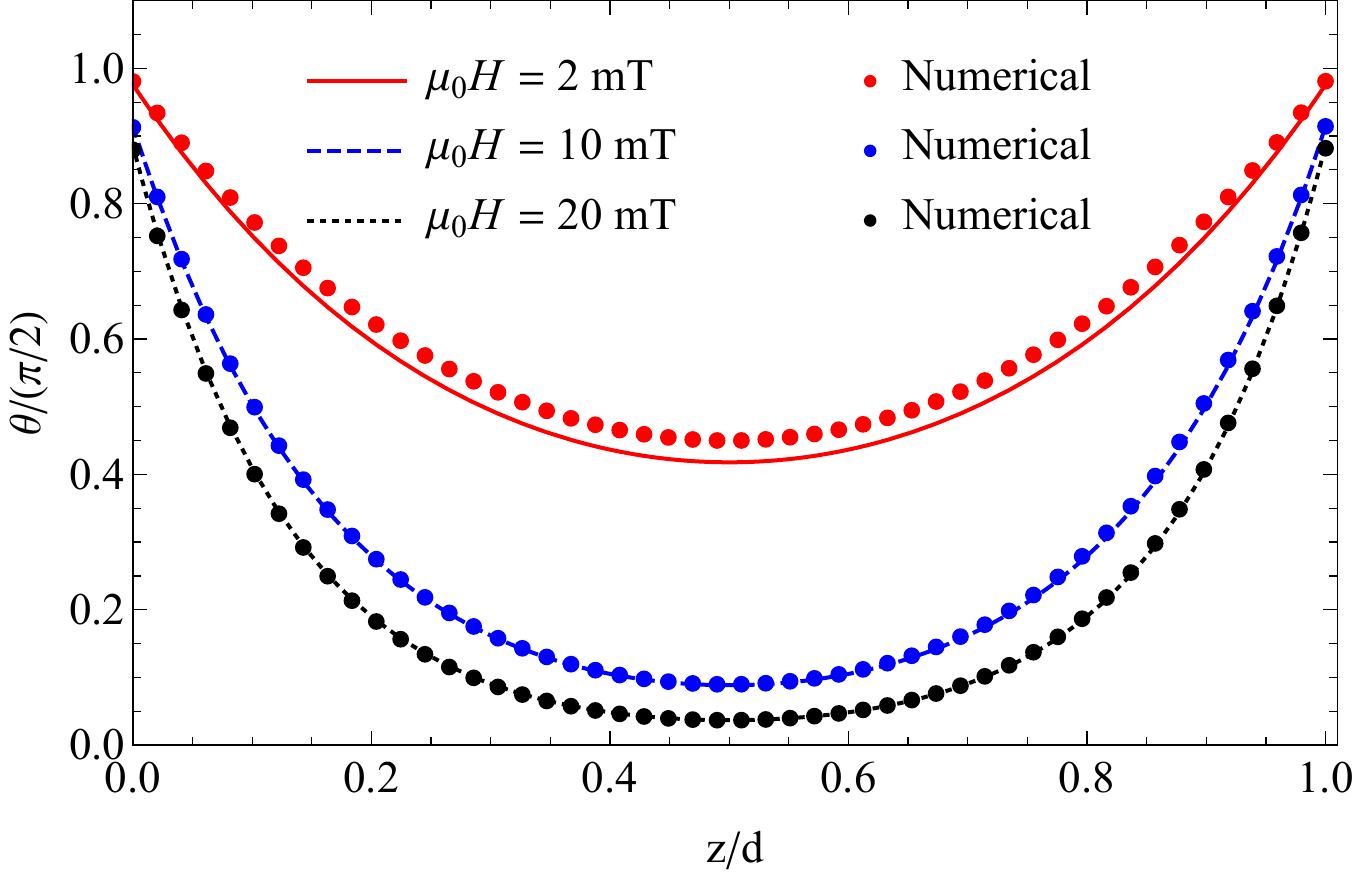}
\caption{(Color online) For large magnetic fields, the numerically calculated polar angle of the director is in agreement with Eq.~(\ref{angles2}).}
\label{Bild4}
\end{figure}

\subsection{Comparison of analytic approximations with numerics}

\noindent
A comparison of analytic and numeric results for the director polar angle 
$\theta(z)$
is made in Figs.~\ref{Bild3} and \ref{Bild4} for small and large magnetic fields, respectively. We find a good agreement for small magnetic fields up to 0.7\,mT and for large magnetic fields above 4\,mT. It should be emphasized that the values of the magnetic fields at which the approximations become valid depend on the values of the static parameters. We use the values Eqs.~(\ref{Wextracted})-(\ref{A1extracted}) extracted from the fits to the macroscopic model.

\begin{figure}[htb]
\includegraphics[width=3.3in]{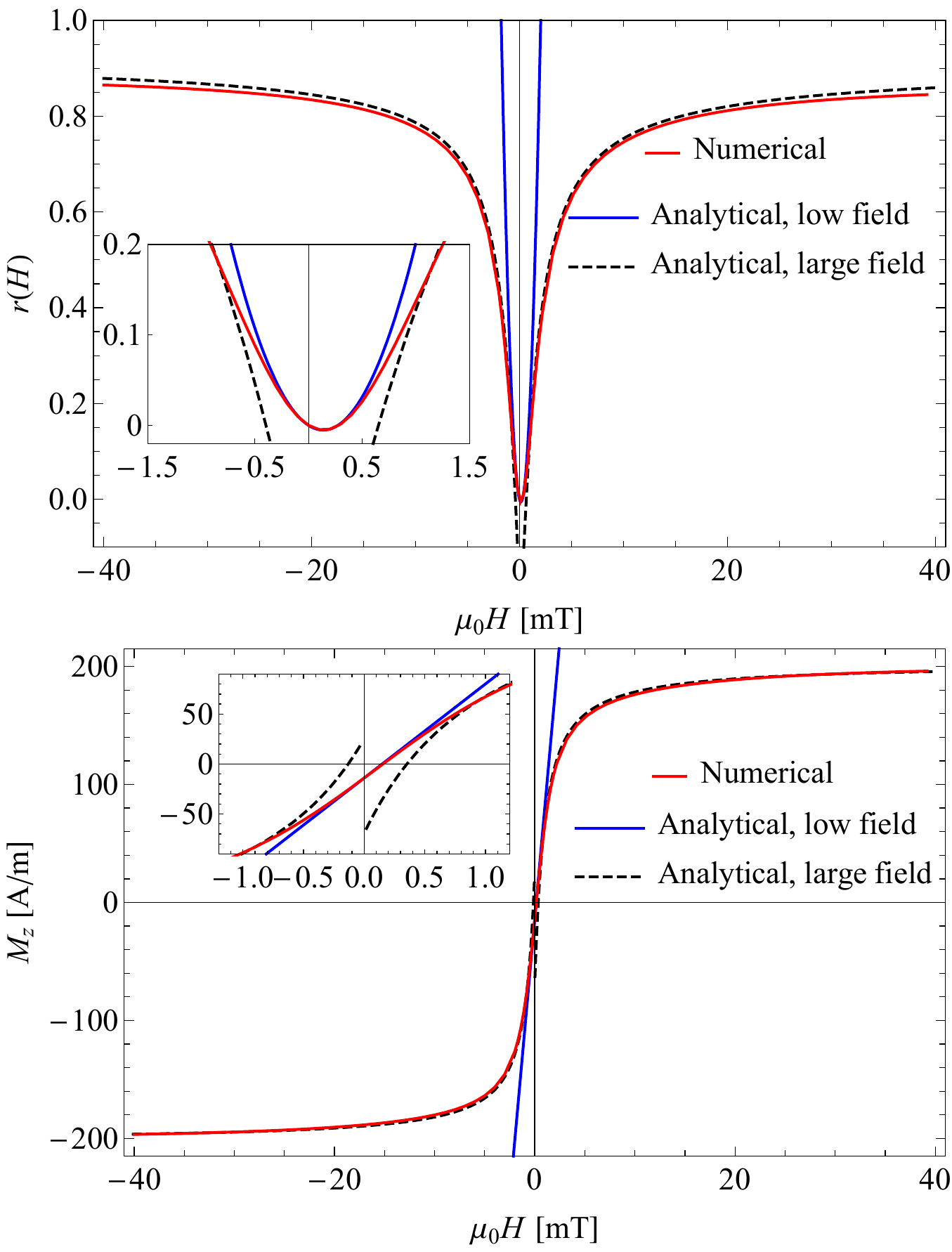}
\caption{(Color online) Comparison of numeric and analytic results at low and high values of the applied magnetic field: (top) magnetic field dependence of the normalized phase difference for small magnetic fields is in agreement with Eq.~(\ref{up1}) below 0.5\,mT, whereas the approximation for large magnetic fields, Eq.~(\ref{phasehigh}), is within one percent of the numerical value already when above 0.8\,mT. 
(bottom) Magnetic field dependence of the 
$z$ component of the magnetization for small magnetic fields is in agreement with Eq.~(\ref{maglow}) below 0.5\,mT, whereas the approximation for large magnetic fields, Eq.~(\ref{maghigh}), is within one percent of the numerical value already when above 0.8\,mT.}
\label{Bild5}
\end{figure}

In Fig.~\ref{Bild5} we compare analytic and numeric results for the 
$z$ component of the magnetization and the normalized phase difference. Again we find a good agreement between the results at similar ranges of the magnetic field. From the insets of Fig.~\ref{Bild5} one can conclude that for our system a magnetic field as small as 1\,mT can be considered as large already.
The notable discrepancy of the numeric and analytic normalized phase difference at large magnetic fields is due to the fact that one has expanded the expression for the phase difference, Eq.~(\ref{eqfaza}), up to the order $\theta^4$. Since $\theta$ does not saturate to zero, this means that the constant term of Eq.~(\ref{phasehigh}) is slightly different from the actual value determined numerically.

The agreement between experimental data and the model for two key static properties
underscores that we have solid ground for the analysis of the dynamic results which now follows.

\section{Switch-on dynamics}
\label{sec:ON}

\noindent
In this Section we present the experimental and theoretical results of the dynamics that takes place when the magnetic field is switched on.

\begin{figure}[htb]
\includegraphics[width=3.3in]{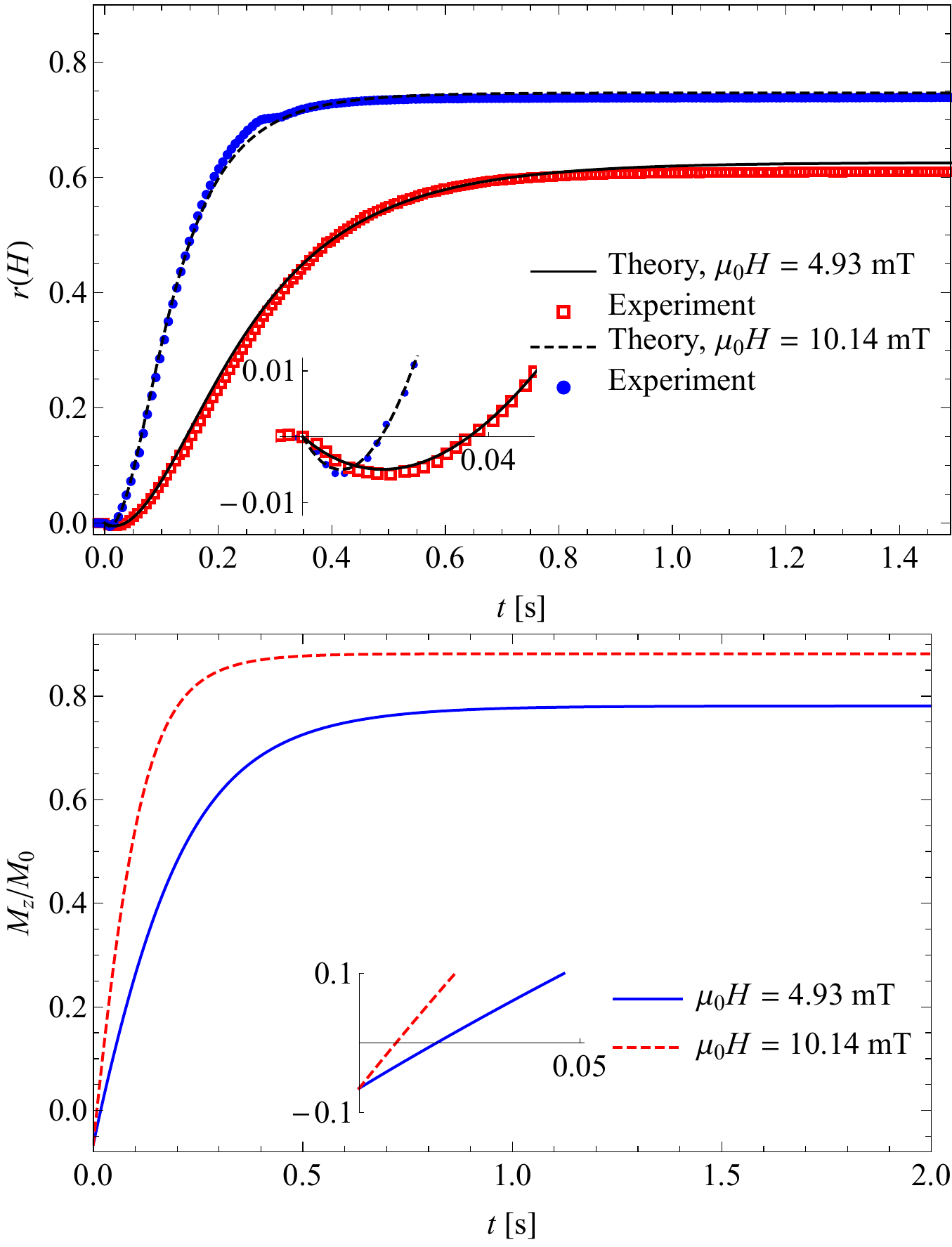}
\caption{
(Color online) 
Top: time evolution of the measured normalized phase difference, $r(H)$, fitted by the dynamic model Eqs.~(\ref{f})-(\ref{bijD}).
The linear-quadratic onset of $r(H)$ is in accord with the analytic result given in Eq.~(\ref{short}). 
Bottom: the corresponding theoretical time evolution of $M_{z}/M_{0}$, initially growing linearly as given in Eq.~(\ref{linear}). 
}
\label{Bild19}
\end{figure}

In Fig.~\ref{Bild19} we plot the comparison of experimental and theoretical data for the dynamics of the normalized phase difference (top) as well as the theoretical results for the normalized $z$ component of the
magnetization (bottom) for two values of the applied magnetic field. 
As an inset we show that for small times the magnetization grows linearly, which is also obtained
analytically in Sec.~\ref{sec-initial}. As expected the rise time for the magnetization
is reduced as the applied magnetic field is increased. 
The inset for the top graph shows that the initial phase difference is quadratic in time, which is again obtained also analytically, Sec.~\ref{sec-initial}.

The fits for the comparison of the experimental and theoretical normalized phase difference are performed by varying the dynamic parameters taking into account the fundamental restrictions \cite{tilenshort} on their values, at fixed values 
of the static parameters Eqs.~(\ref{Wextracted})-(\ref{A1extracted}).
The model captures the dynamics very well for all times from the onset to the saturation. 
The extracted values of the dynamic parameters are
\begin{eqnarray}
	\gamma_1 &\sim& 0.13\,{\rm Pa\,s},\label{labgam}\\
	b_\perp^D &\sim& 1.5\times 10^5\,{\rm Am/Vs^2},\label{labbperp}\\
	\chi_2^D &\sim& 4\,({\rm Pa\,s})^{-1}.
\end{eqnarray}
The dissipative cross-coupling coefficient $\chi_2^D$ is within the allowed interval determined by the restriction \cite{tilenshort}
\begin{equation}
|\chi_2^D|<\sqrt{\frac{b_\perp^D}{\gamma_1M_0^2}}\approx 5.4\,({\rm Pa\,s})^{-1}.
\end{equation}
The remaining two dynamic parameters do not affect the dynamics significantly
and are set to $b_\parallel=b_\perp$ and $\chi_1^D=0$.

\begin{figure}[htb]
\includegraphics[width=3.4in]{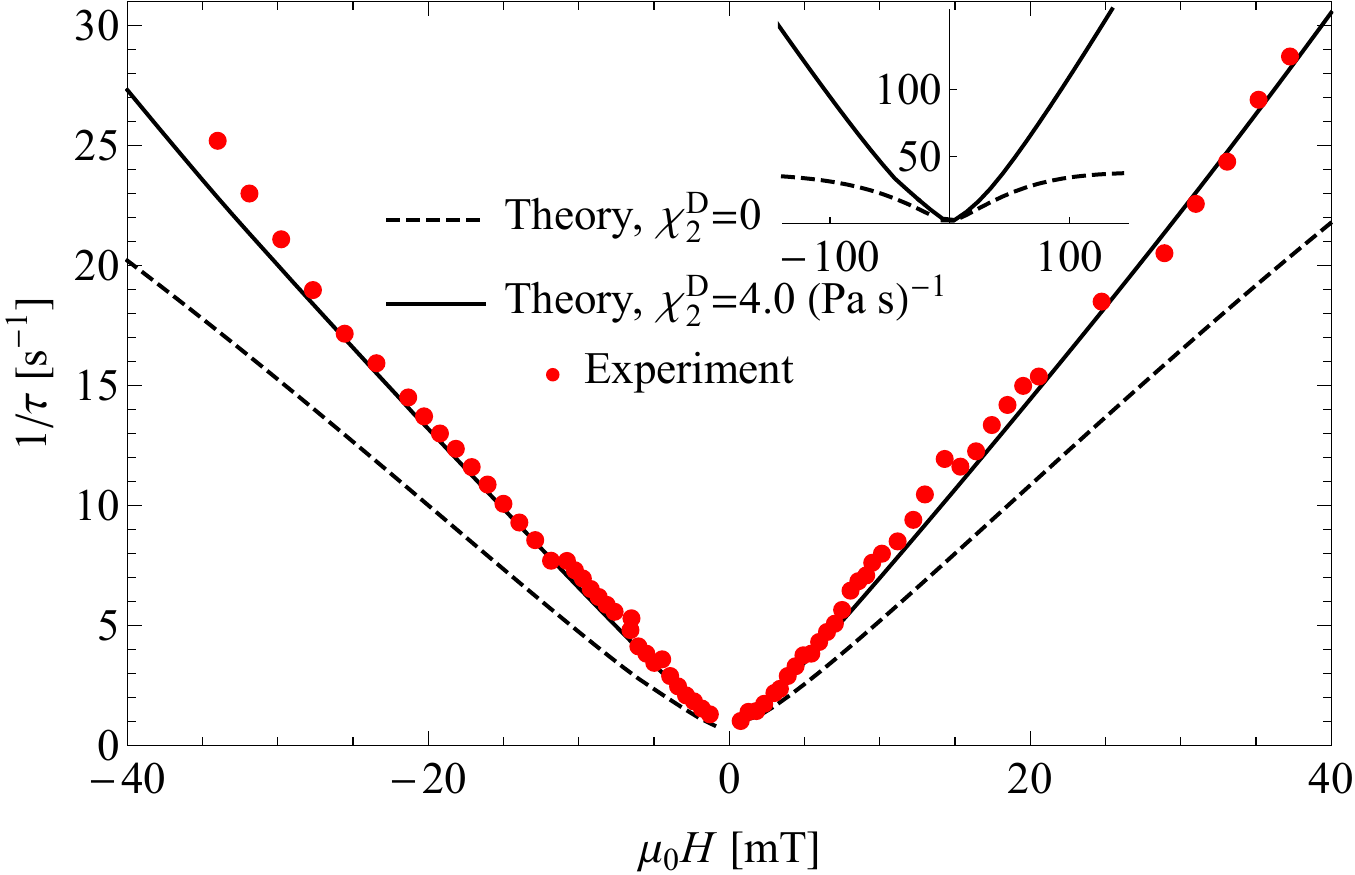}
\caption{(Color online) 
The overall relaxation rate, $1/\tau(H)$, as a function of the magnetic field $\mu_0H$, extracted from
the experimental data and the theoretical results using the fitting function Eq.~(\ref{sigmoidal}).
Inset: without the dynamic cross-coupling, the relaxation rate levels off already at low fields (dashed).}
\label{Bild22}
\end{figure}

To extract from the time evolution of the normalized phase difference, Fig.~\ref{Bild19} (top), a switching time $\tau$ as a measure of an overall relaxation rate of the dynamics, we use a squared sigmoidal model function 
\begin{equation} 
	f(t) = {C'}\left[1-{1+C\over 1+C\exp(-2t/\tau)}\right]^2.
	\label{sigmoidal}
\end{equation} 
Remarkably, the relaxation rate, $1/\tau(H)$, shows a linear dependence on $H$, Fig.~\ref{Bild22}.  
We were first interested in the effect of the dissipative cross coupling on $1/\tau(H)$. 
We find that a reasonably strong dynamic cross coupling $\chi_2^D$ is needed in order to obtain the observed linear magnetic field dependence of the relaxation rate. In the absence of this dynamic cross coupling, Fig.~\ref{Bild22}, the relaxation rate levels off already at low fields as expected since the transient angle between $\bf M$ and $\bf n$ gets larger,
and starts to decrease for even higher magnetic fields.



The best match of the relaxation rates $1/\tau(H)$ extracted from the experimental 
data and the model, Fig.~\ref{Bild22}, allows for a robust evaluation of the dissipative 
cross-coupling between the magnetization and the director: 
\begin{equation}
	\chi_2^D = (4.0\pm 0.7)\,({\rm Pa\,s})^{-1}.
\end{equation}

\subsection{Initial dynamics}
\label{sec-initial}

\noindent
We investigate the initial dynamics of the normalized phase difference and magnetization upon application of the magnetic field. Up to linear order we also take into account the pretilt.
Initially, $\mathbf{n}$ and $\mathbf{M}$ are parallel to $\mathbf{n}_S$. 
%
Keeping the modulus of the magnetization exactly fixed, the initial thermodynamic forces Eqs.~(\ref{h^M}) and (\ref{h^n}) are
\begin{equation}
\begin{split}
\label{inforco}
&\mathbf{h}^n=0,\\
&\mathbf{h}^{\perp M}=\mu_0H(\varphi_s,0,-1).
\end{split}
\end{equation}
where $\mathbf{h}^{\perp M}$ is the projection of $\mathbf{h}^M$ perpendicular to $\bf M$.
With that, the initial quasi-currents are 
\begin{eqnarray}
	Y_i&=&\chi_{ij}^Dh_j^{\perp M}+\chi^R\epsilon_{ijk}n_jh^{\perp M}_k\quad \Rightarrow\nonumber \\
{\bf Y}&=&\mu_0H(\chi_2^DM_0 \varphi_s,\chi^R,-\chi_2^DM_0),\label{intokoY}\\
X_i&=&b_{ij}^Dh_j^{\perp M}+b_{ij}^Rh_j^{\perp M} \quad\Rightarrow\nonumber \\
{\bf X}&=&\mu_0H(b_\perp^D\varphi_s,-(b_1^R+b_2^R)M_0,-b_\perp^D).\label{intokoX}
\end{eqnarray}
At finite $\chi_{2}^D$ and zero $\chi^R$ it follows from Eq.~(\ref{intokoY}) that the $z$ component of the director field responds linearly in time as well as linearly in the magnetic field for small times:
\begin{equation}
	n_z(t)\approx \varphi_s + \chi_{2}^{D}M_0\mu_0 H\, t.
	\label{nzlinear}
\end{equation}

As a contrast, if $\chi_{2}^D$ is zero, the director responds through the nonzero molecular field $h_z^n$ due to the static coupling $A_1$,
\begin{equation}
\label{nzquadratic'}
h_z^n=-A_1M_0 M_z(t) = -A_1M_0b_\perp^D\mu_0H\, t,
\end{equation}
where $M_z(t) = b_\perp^D\mu_0H t$ is the initial response of the $z$ component of the magnetization, Eq.~(\ref{intokoX}). 
The $z$ component of the director field thus responds quadratically in time rather than linearly,
\begin{equation}
	n_z(t)\approx \varphi_s + \frac{A_1M_0b_\perp^D\mu_0H}{2\gamma_1}t^2.
	\label{nzquadratic}
\end{equation}

\begin{figure}[htb]
	\includegraphics[width=3.3in]{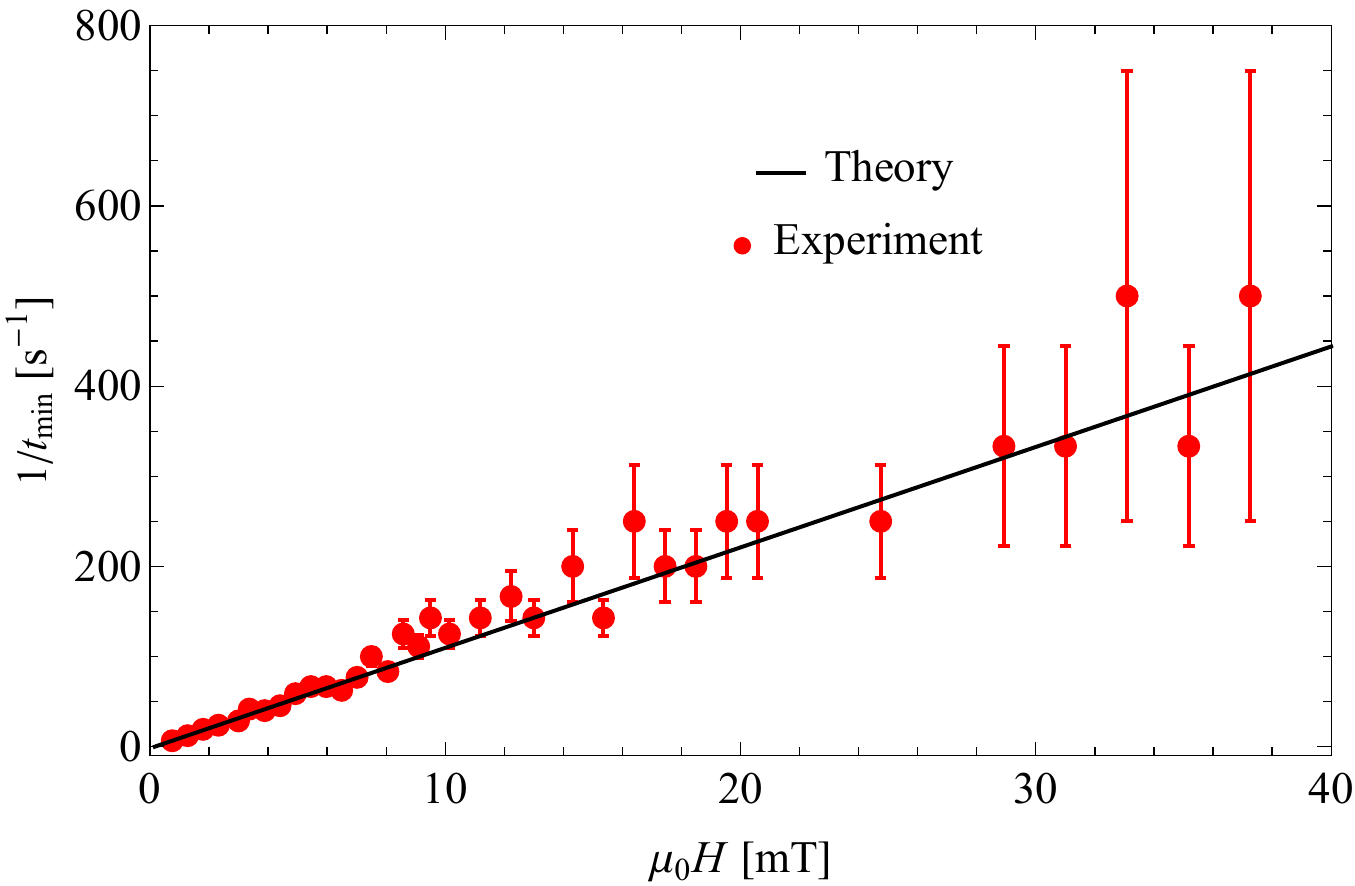}
	\caption{
	(Color online) 
	Inverse of the time of the minimum determined from the measured normalized phase difference as a function of the magnetic field. The linear behavior in magnetic field is in agreement with Eq.~(\ref{tmin}).}
	\label{Bild17}
\end{figure}

For small times $t$ we can express the refractive index Eq.~(\ref{refract}) as
\begin{equation}
	n_e(t)\approx n_{e0}\left[1-\frac{n_{e0}^2-n_o^2}{2n_o^2}\left(\varphi_s + \chi_2^DM_0\mu_0Ht\right)^2\right].
\end{equation}
The coefficients $a$ and $b$ from Eq.~(\ref{ab}) are then
\begin{equation}
\begin{split}
\label{abiniti}
&a\approx \frac{k_0d}{2}\left[n_e(t)-n_o\right]\left[1-2\left(\chi^R\mu_0H\right)^2t^2\right],\\
&b\approx \frac{k_0d}{2}\left[n_e(t)-n_o\right]\left(-2\chi^R\mu_0H\right)t\\
\end{split}
\end{equation}
%
and the normalized intensity of the transmitted light for small times is
\begin{equation}
\begin{split}
\label{intrev}
&\frac{I}{I_0}\approx \sin^2\left(\frac{\phi_0}{2}\right)-r_0\varphi_s\chi_2^D\mu_0HM_0\phi_0\sin(\phi_0)t\\ 
&-\left[\frac{r_0}{2}(\chi_2^D\mu_0HM_0)^2\phi_0\sin(\phi_0)+4(\chi^R\mu_0H)^2\sin^2\left(\frac{\phi_0}{2}\right)\right]t^2.
\end{split}
\end{equation}
In the lowest order of $t$, for the phase difference one gets a linear term that is also linear in pretilt and a quadratic term which does not vanish if the pretilt is zero:
\begin{eqnarray}
\label{short}
	r(H)&\approx &r_0\left[\left(\chi_2^DM_0\mu_0H\right)^2t^2+2\varphi_s \chi_2^DM_0\mu_0H t\right]\nonumber \\
	&\equiv& k^2t^2+pt.
\end{eqnarray}
Eq.~(\ref{short}) will be used to extract the dissipative cross coupling coefficient $\chi_2^D$ and the pretilt $\varphi_s$ from the experimental data. Furthermore, from Eq.~(\ref{short}) one can see that in the case of positive (negative) pretilt the normalized phase difference has a minimum at negative (positive) magnetic fields. 
By measuring the time of this minimum, Fig.~\ref{Bild17}, 
\begin{equation}
	\label{tmin} 
	t_{\mathrm{min}}=-\frac{\varphi_s}{\chi_2^D\mu_0HM_0},
\end{equation}
one can calculate the ratio of the pretilt and the dissipative cross coupling.
If $\chi_2^D=0$, the time of the minimum decreases more slowly with increasing magnetic field:
\begin{equation}
	t_{\mathrm{min}}=\sqrt{-\frac{2\gamma_1\varphi_s}{A_1 b_\perp^D\mu_0HM_0}}.
\end{equation}

The normalized phase difference evaluated at $t_{\mathrm{min}}$ is of second order in the pretilt:
\begin{equation}
	\label{eqpret}
	r(H)_{\mathrm{min}}=-r_0\varphi_s^2.
\end{equation}
The minimum value Eq.~(\ref{eqpret}) is independent of the applied magnetic field. This can be explained by the fact that the director field goes through an intermediate state which is approximately aligned with the glass plates of the cell. 
\begin{figure}[htb]
	\includegraphics[width=3.3in]{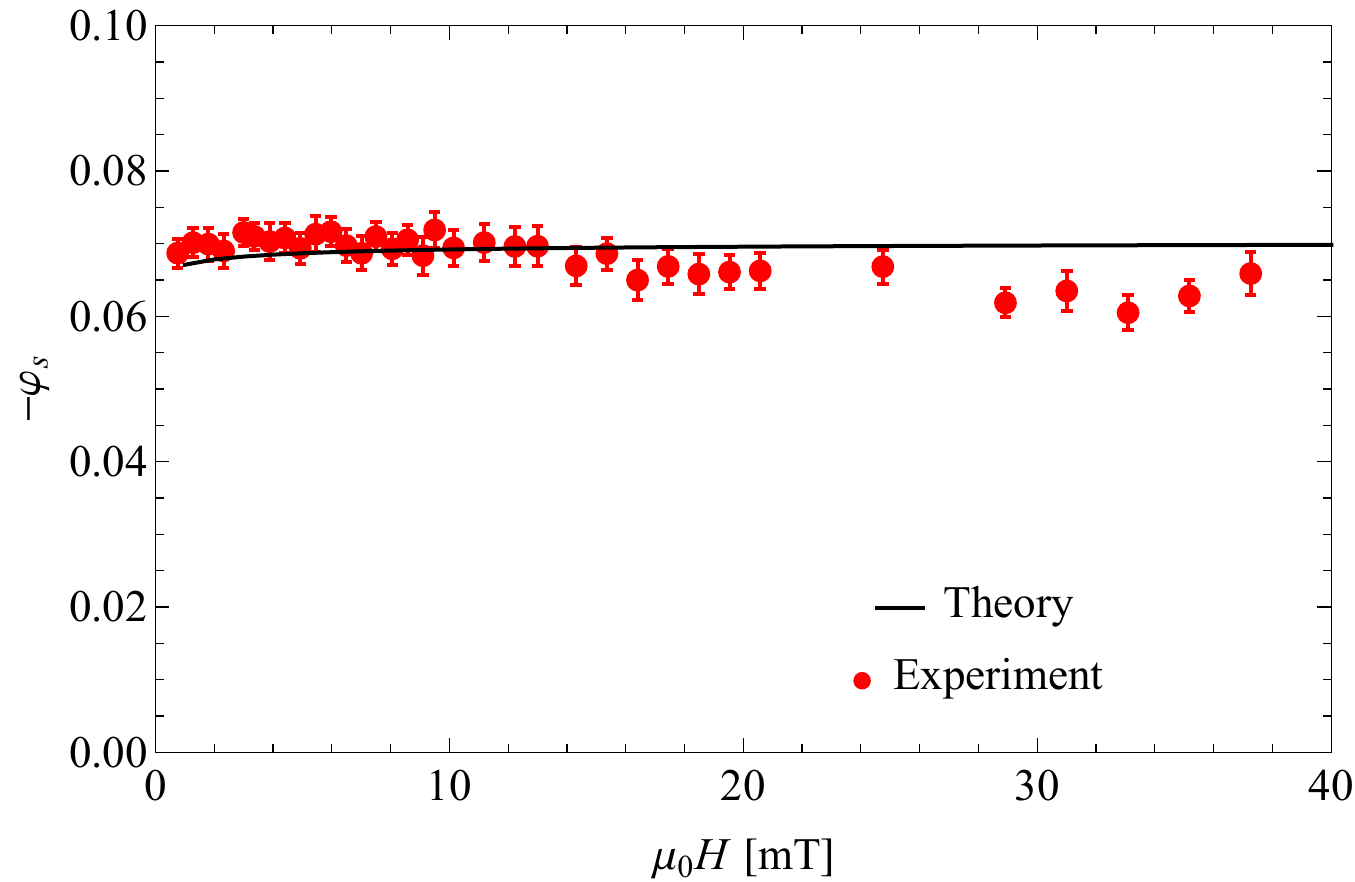}
	\caption{
	(Color online) 
	Pretilt, determined from experimental data using Eq.~(\ref{eqpret}).}
	\label{Bild18}
\end{figure} 

We note that if both the dissipative cross coupling coefficient $\chi_2^D$ and the pretilt $\varphi_s$ are zero, the normalized phase difference initially grows as $t^4$. 
%


Assuming a negative pretilt, Eq.~(\ref{short}) predicts a minimum for positive magnetic fields, which is also seen in experiments, Fig.~\ref{Bild19} (top). 
In Fig.~\ref{Bild17} we show experimental inverse times of the minima. The large error at high magnetic fields is due to the time resolution limitations (1\,ms).
>From the linear behavior predicted by Eq.~(\ref{tmin}) we can extract the ratio between the dissipative cross-coupling and the pretilt. Independently we can extract the pretilt by measuring the values of the minima, Fig~\ref{Bild18}. 

\begin{figure}[htb]
	\includegraphics[width=3.3in]{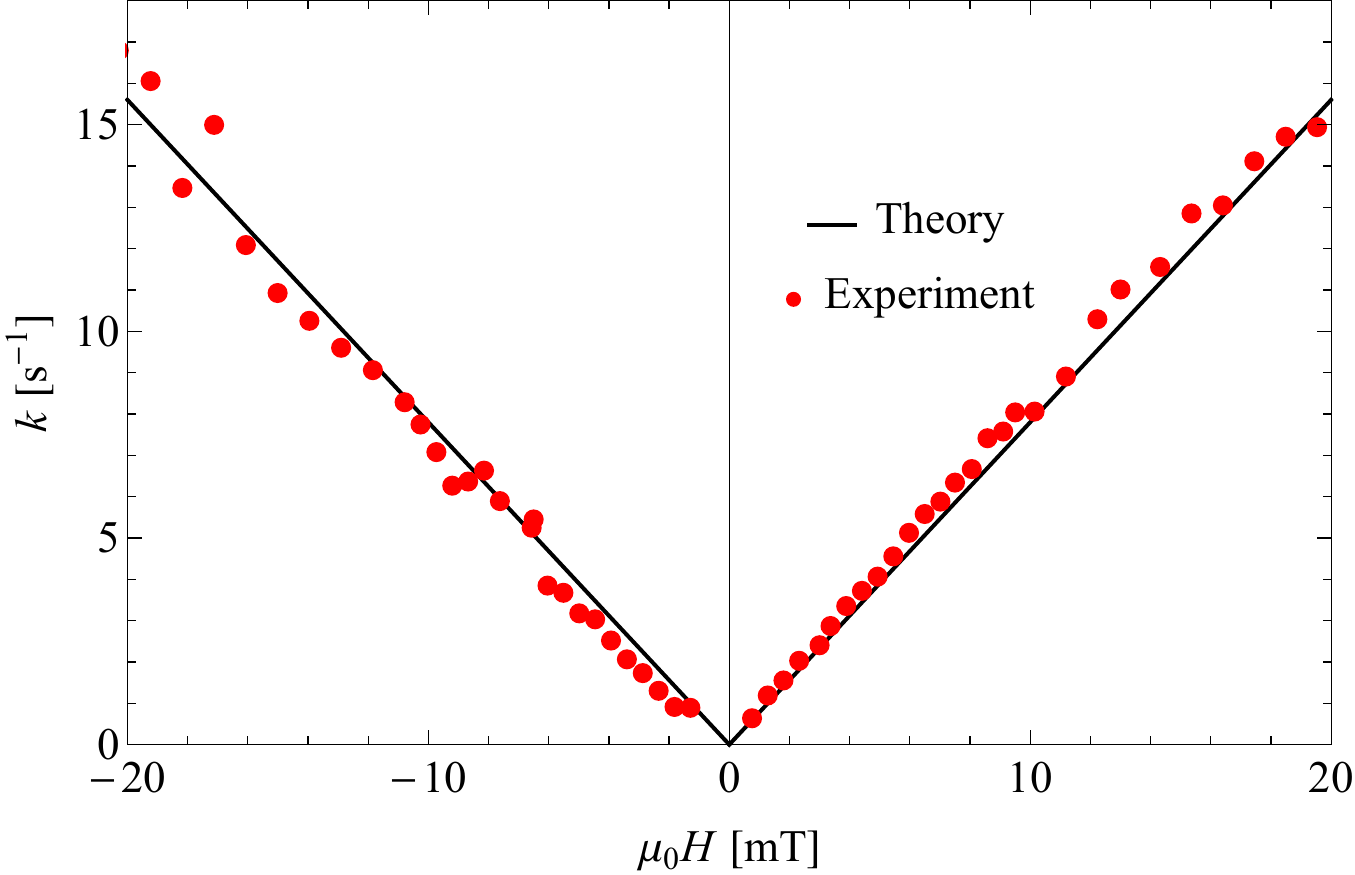}
	\caption{
	(Color online) 
	The coefficient $k$ of Eq.~(\ref{short}) as a function of the magnetic field $\mu_0H$. The straight line fits are used to extract $\chi_2^D$.}
	\label{Bild20}
\end{figure}

\begin{figure}[htb]
	\includegraphics[width=3.3in]{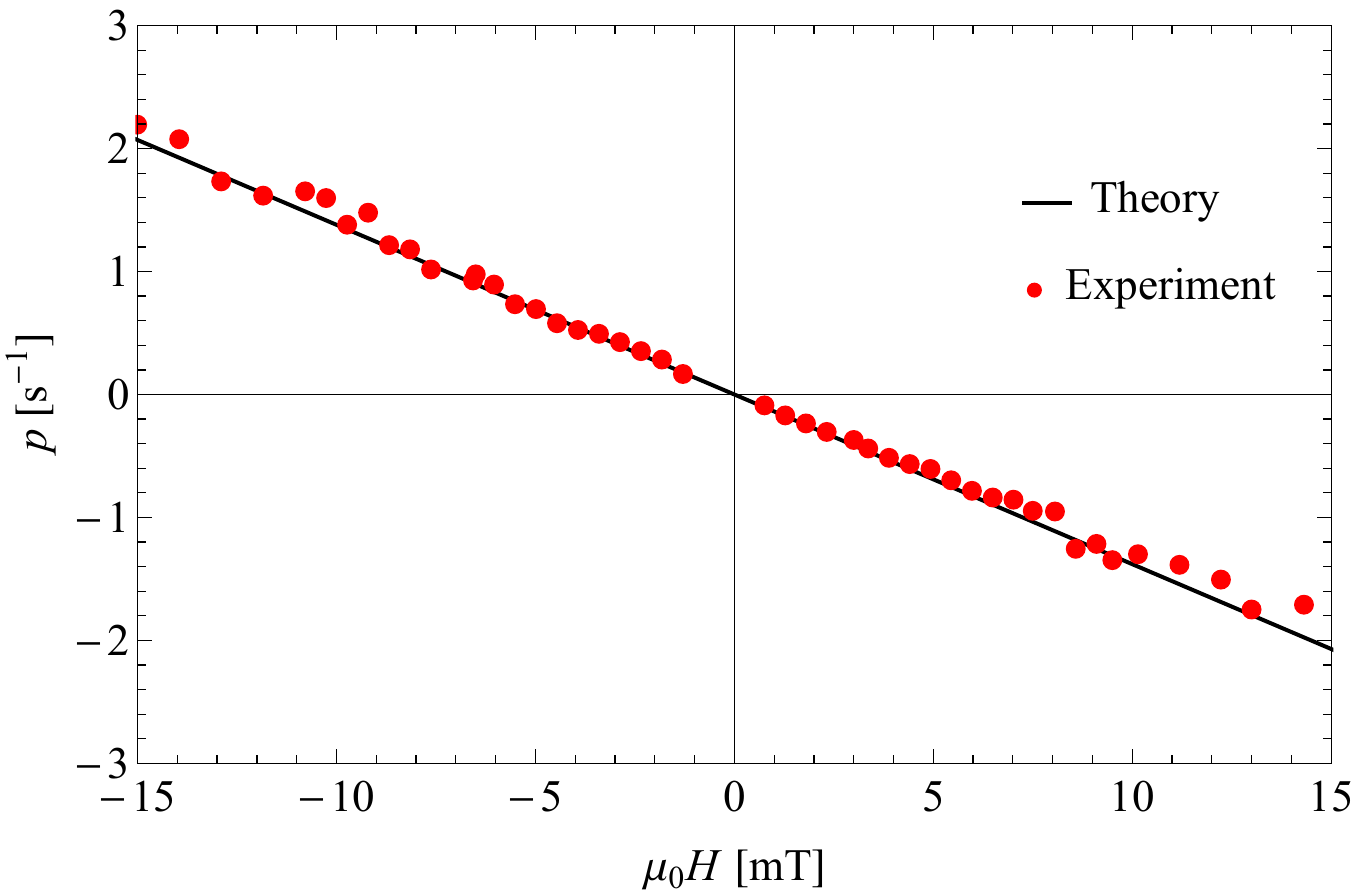}
	\caption{
	(Color online) 
	The coefficient $p$ of Eq.~(\ref{short}) as a function of the magnetic field $\mu_0H$. The straight line fit is used to extract $\varphi_s$.}
	\label{Bild21}
\end{figure}

%
Fitting Eq.~(\ref{short}) to the initial time evolution of measured normalized phase differences (like those presented in Fig.~\ref{Bild19}) for several values of the magnetic field $\mu_0H$, 
we determine the parameters $k$ and $p$ shown in Fig.~\ref{Bild20} and Fig.~\ref{Bild21}, respectively.
Therefrom we extract the value of the dissipative cross-coupling parameter $\chi_2^D$ between director and magnetization,
\begin{equation}
	\chi_2^D \sim (4.0\pm 0.5)\,({\rm Pa\,s})^{-1},
\end{equation}
and from the parameter $p$ of Eq.~(\ref{short}) we extract the pretilt, 
\begin{equation}
	\varphi_s \sim -0.065 \pm 0.01. 	
\end{equation}
%

The normalized $z$ component of the magnetization Eq.~(\ref{eq0}) is linear in $t$:
\begin{equation}
	\label{linear}
	\frac{M_z}{M_0}=\varphi_s +  \frac{b_{\perp}^D}{M_0}\mu_0Ht, 
\end{equation}  
which is in accord with Fig.~\ref{Bild19} (bottom).
>From the initial behavior one can therefore directly determine the dissipative coefficient $b_\perp^D$.


Let us define the initial rate of the director reorientation as the time derivative of the director $z$ component at $t=0$,
\begin{equation}
	\frac{1}{\tau_{d}}=\left .\frac{\partial n_z}{\partial t}\right |_{t=0}.
\end{equation} 
For a nonzero dissipative cross-coupling coefficient $\chi_2^D$ the initial rate, Eq.~(\ref{nzlinear}), is
\begin{equation}
	\label{dinamre}
	\frac{1}{\tau_{d}}=\chi_{2}^{D}M_0\mu_0|H|.
\end{equation}
However if $\chi_2^D=0$, the initial rate of the director reorientation is proportional to the $z$ component of the magnetization, Eq.~(\ref{nzquadratic'}),
\begin{equation}
	\label{staticre}
	\frac{1}{\tau_{s}}=\frac{A_1M_0}{\gamma_1}|M_z(t)|.
\end{equation}
The relaxation rates Eqs.~(\ref{dinamre}) and (\ref{staticre}) describe two different mechanisms of the director reorientation. The former is associated with the dynamic coupling of the director and the magnetization, whereas the latter is governed by the static coupling $A_1$ of the director and the magnetization. Here a deviation of the magnetization from the director is needed to exert a torque on the director.

\subsection{Dissipative cross-coupling}

\begin{figure}[htb]
	\includegraphics[width=3.3in]{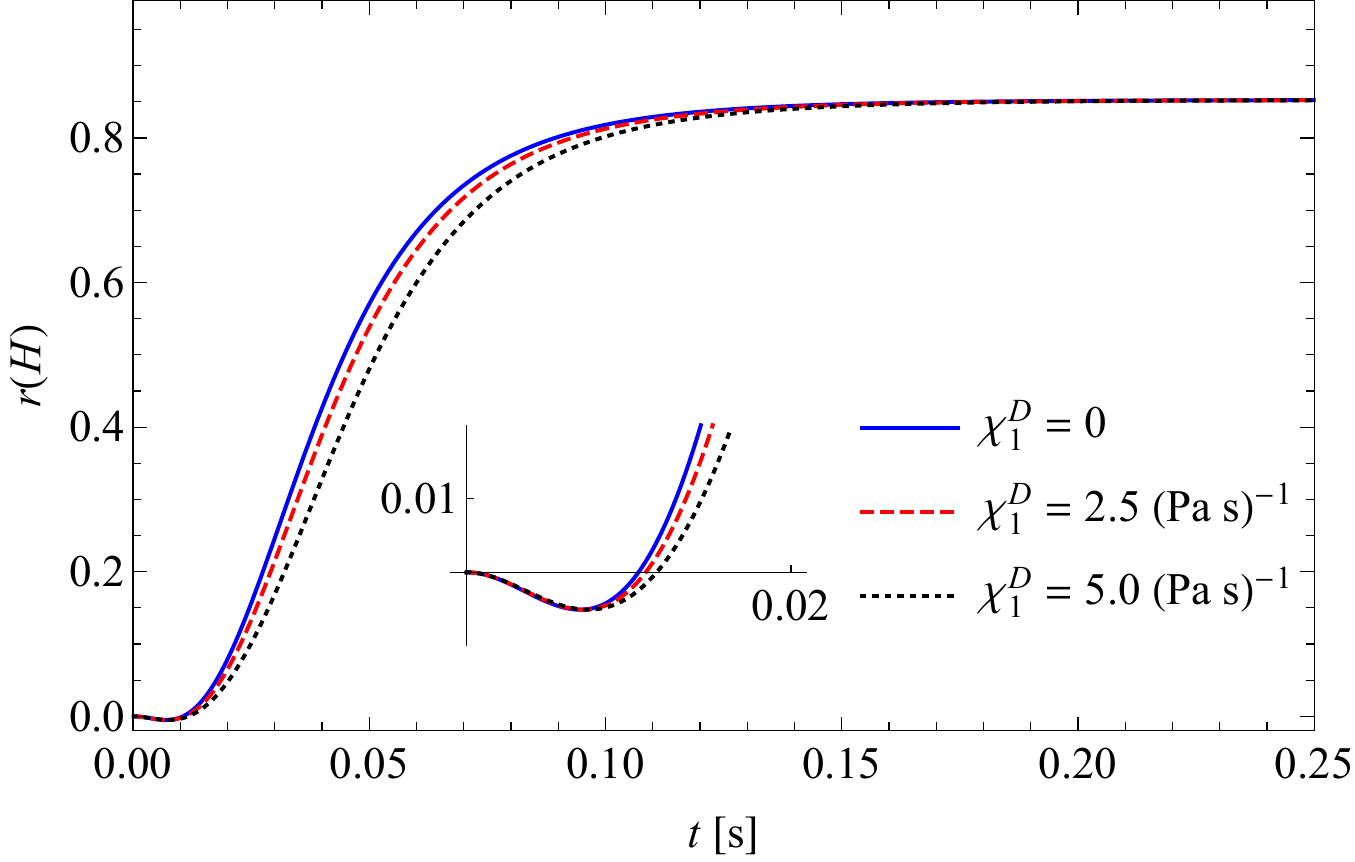}
	\caption{Normalized phase difference at different values of the dissipative cross-coupling parameter $\chi_{1}^D$, $\chi_2^D=0$,  $\mu_0 H=50$\,mT. Inset: the initial behavior is not affected.} 
	\label{Bild9}
\end{figure}

\noindent
We have demonstrated that the dissipative cross-coupling of the director and the magnetization, i.e., the $\chi_{ij}^D$ terms of Eqs.~(\ref{X}) and (\ref{Y}), affects the dynamics decisively and is crucial to explain the experimental results.
It is described by the parameters $\chi_1^D$ and $\chi_2^D$ of Eq.~(\ref{chiD}).
Here we check the sensitivity of the dynamics to the values of these two parameters.
Varying $\chi_1^D$ while keeping $\chi_2^D=0$, Fig.~\ref{Bild9}, we see that the influence of $\chi_1^D$ is rather small and is not substantial.
Moreover, the initial dynamics is not affected, Fig.~\ref{Bild9} (inset). 

\begin{figure}[htb]
	\includegraphics[width=3.3in]{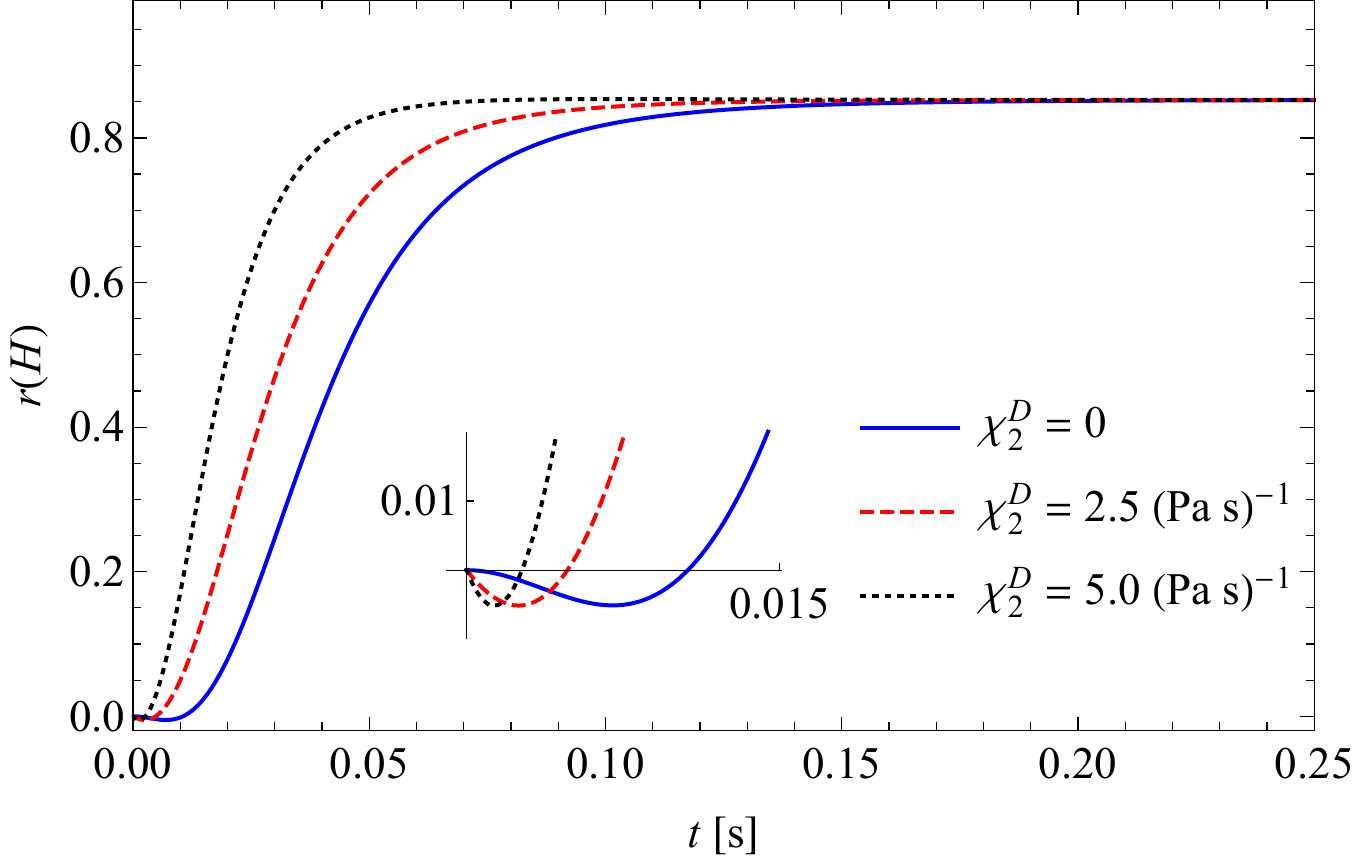}
	\caption{Normalized phase difference at different values of the dissipative cross-coupling parameter $\chi_{2}^D$, $\chi_1^D=0$,  $\mu_0 H=50$\,mT. Inset: the initial behavior is strongly affected as well.} 
	\label{Bild10}
\end{figure}

On the other hand, increasing $\chi_2^D$ strongly reduces the rise time of the normalized phase difference, Fig.~\ref{Bild10}, and also strongly affects the initial behavior (inset). For large values of $\chi_2^D$ one also observes an overshoot in the normalized phase difference.

By inspecting Eq.~(\ref{chiD}) one sees that the influence of $\chi_1^D$ is largest when ${\bf M}\perp{\bf n}$, ${\bf h}^n\parallel{\bf M}$ and ${\bf h}^M\parallel{\bf n}$.
On the other hand, the influence of $\chi_2^D$ is largest when ${\bf M}\parallel{\bf n}$.
Since $\bf M$ and $\bf n$ are initially parallel and moreover the transient angle between them never gets large due to the strong static coupling compared to the magnetic fields applied, it is understandable that $\chi_2^D$ affects the dynamics more than $\chi_1^D$.

\subsection{Reversible cross-coupling}
\label{sec:reversible}

\noindent
The reversible cross-coupling of the director and the magnetization, described by the $\chi^R$ terms of Eqs.~(\ref{XR}) and (\ref{YR}), has not been considered up to this point.
We focus on the reversible cross coupling coefficient $\chi^R$ and put both reversible tensors $b_{ij}^R$ and $(\gamma^{-1})_{ij}^R$ of Eqs.~(\ref{b_ij^R}) and (\ref{gamma_inv}) to zero.

\begin{figure}[htb]
	\includegraphics[width=3.3in]{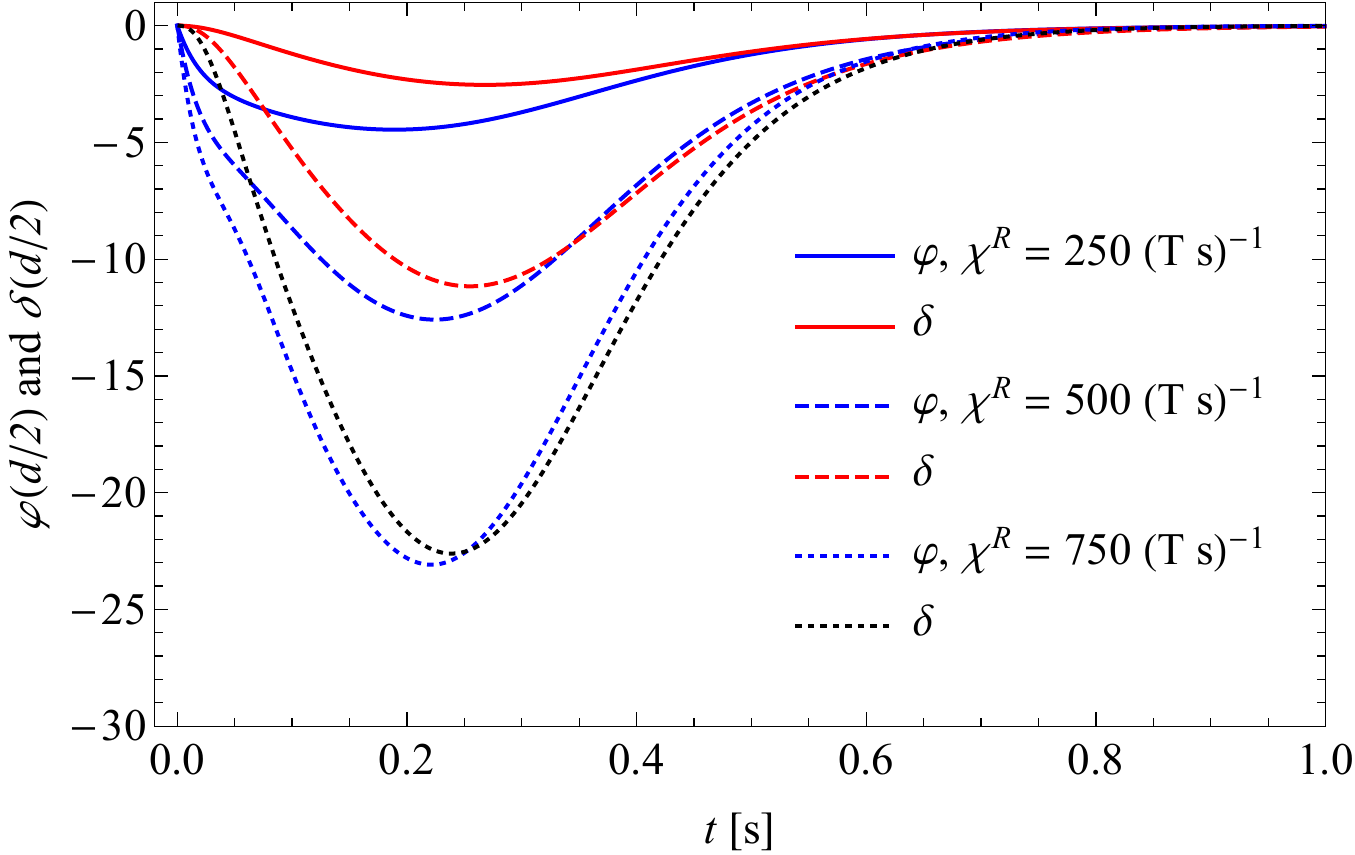}
	\caption{
	(Color online)
	The time dependences of the azimuthal angles (degrees) of the director ($\varphi$) and the magnetization ($\delta$) at $z=d/2$ for different values of $\chi^R$, $\chi_1^D=\chi_2^D=0$, $\mu_0H=10$\,mT.}
	\label{Bild15}
\end{figure}

If the reversible currents are included, both variables wander out of the $xz$ plane dynamically, which will be described by the azimuthal angles $\delta$ and $\varphi$ of the magnetization and the director, respectively, defined by
$\mathbf{M}=M_0(\cos\delta\sin\psi,\sin\delta\sin\psi,\cos\psi)$, $\mathbf{n}=(\cos\varphi\sin\theta,\sin\varphi\sin\theta,\cos\theta)$. 
The dynamic behavior of both azimuthal angles is shown in Fig.~\ref{Bild15}.

Contrary to the polar angles 
we find that the response of the azimuthal angle of the director is faster than that of the magnetization.
>From Fig.~\ref{Bild15} we read off that 
the maximum azimuthal angles increase with $\chi^R$, being higher for the magnetization than for the director.

We note again that here we only included the reversible cross-coupling $\chi^R$. 
>From the initial quasi-currents Eqs.~(\ref{intokoY}) and (\ref{intokoX}) one can see that the initial azimuthal response of the magnetization can be faster than that of the director if the coefficients of the tensor $b_{ij}^R$ are sufficiently large,
\begin{equation}
	|b_1^R+b_2^R|>|\chi^R|/M_0.
\end{equation}

%


There exists a direct way of detecting the possible dynamics in the $xy$ plane. The intensity of the transmitted light in the experiments with crossed polarizers at $45^\circ$ and $-45^\circ$ is given by Eq.~(\ref{normint}),
\begin{equation}
\frac{I}{I_0}=\frac{a^2}{c^2}\sin^2(c).
\end{equation}
It is this quantity that is typically measured.
On the other hand, crossed polarizers at $0^\circ$ and $90^\circ$ give us the intensity
\begin{equation}
	\label{intxy}
	\frac{I}{I_0}=\frac{b^2}{c^2}\sin^2(c),
\end{equation}
with a and b given by Eq.~(\ref{ab}). This method is better suited for detecting the $xy$ dynamics, since $b$ is more sensitive to the deviation of the director field from the $xz$ plane.

%

\begin{figure}[htb]
	\includegraphics[width=3.3in]{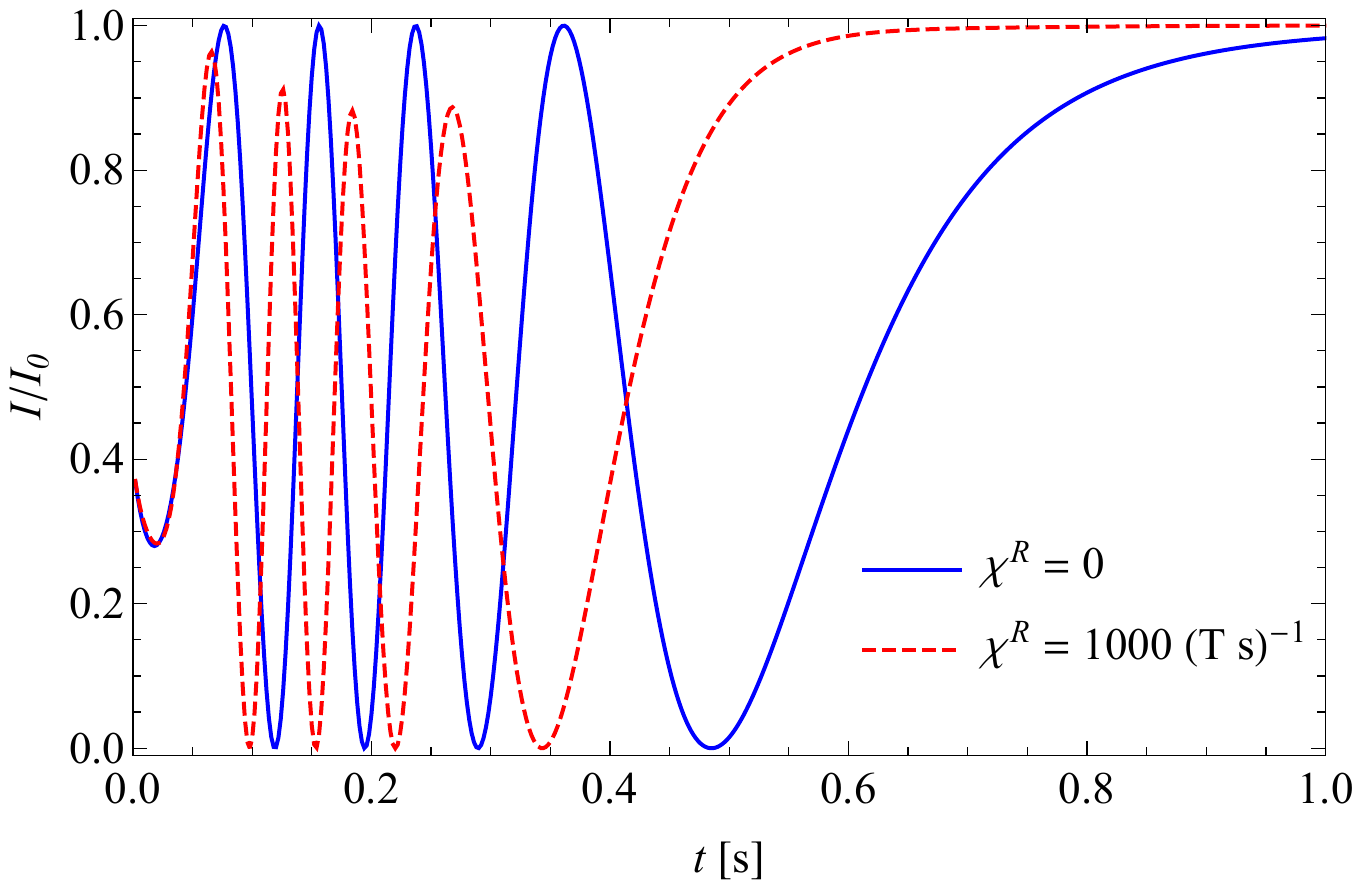}
	\caption{
	(Color online) 
	Time dependence of the normalized intensity of transmitted light for zero and nonzero values of the reversible cross coupling coefficient $\chi^R$; $\mu_0H=5$\,mT.}
	\label{Bild16}
\end{figure}

Our numerical calculations have revealed that, owing to the reversible dynamics, the magnetization and the director are not confined to the $xz$ plane. 
As a consequence, the maxima of the time-dependent intensity of transmitted light are lower than unity, Fig.~\ref{Bild16}, in contrast to the case of a purely in-plane (dissipative) dynamics. Observation of the lower maxima could thus be an indication of the azimuthal dynamics. This effect is more prominent at higher magnetic fields and at higher values of the reversible cross coupling coefficients.

In recent experiments no clear-cut consequences of the azimuthal dynamics have been found using crossed polarizers at $0^\circ$ and $90^\circ$. 
In the following we will therefore discard the reversible dynamics.

\section{Switch-off dynamics}
\label{sec:OFF}

\noindent
Dynamics of the normalized phase difference after switching off the magnetic field has been also measured. In experiments, the initial state is obtained by switching on the desired magnetic field and waiting for a couple of seconds. Contrary to the previous experiments, here the initial state is not homogeneous.

\begin{figure}[htb]
	\includegraphics[width=3.3in]{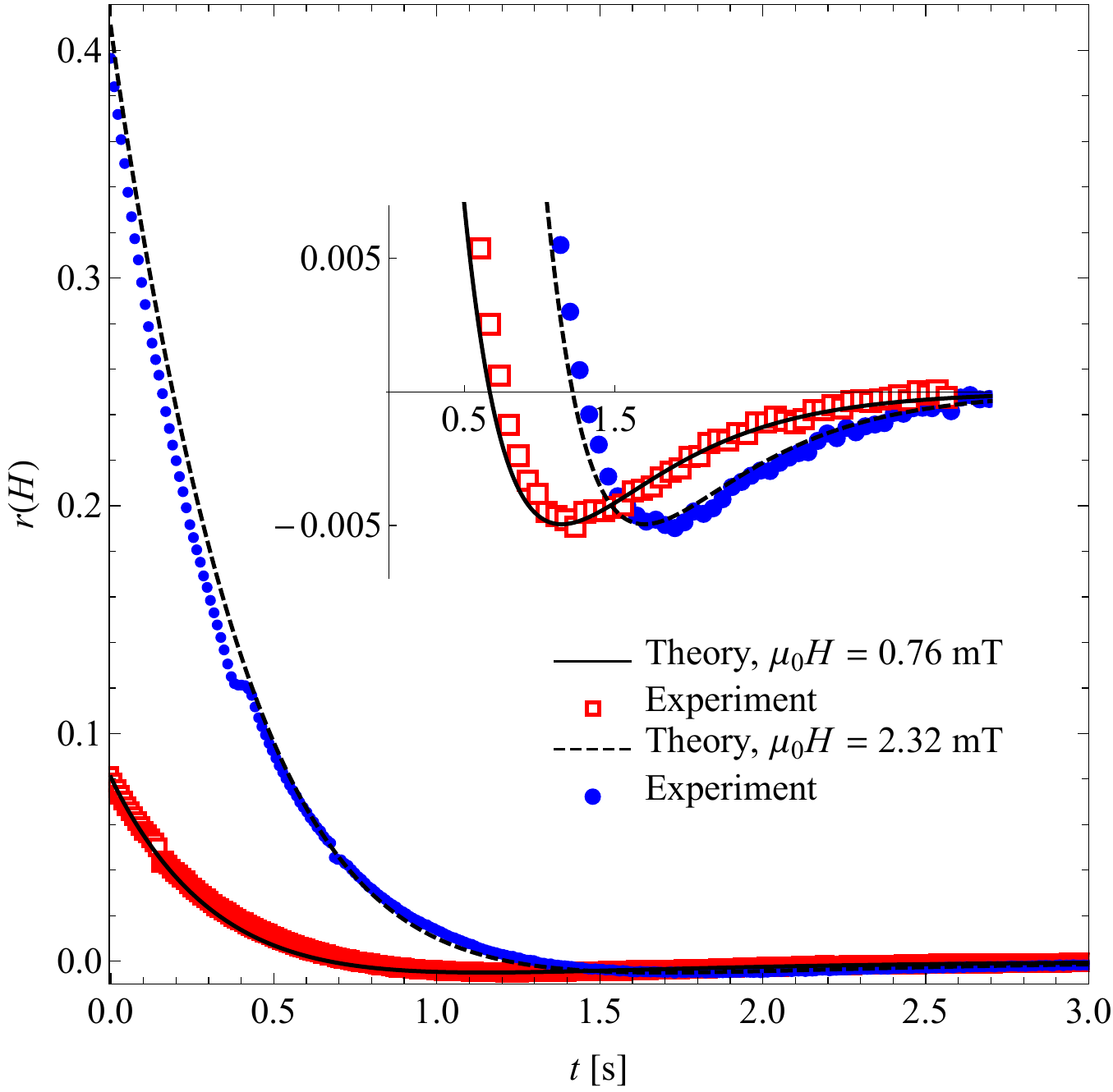}
	\caption{ (Color online) Experimental and numerical normalized phase difference as a function of time at different values of the applied magnetic field.}
	\label{Bild25}
\end{figure}

In Fig.~\ref{Bild25} we compare the experimental and numerical normalized phase difference at two different fields. We observe, similarly to the switch-on case, that the normalized phase difference goes through a minimum. This is again explained by the fact that the director field goes through a state, which is approximately aligned with the surfaces of the glass plates. 
\begin{figure}[htb]
	\includegraphics[width=3.3in]{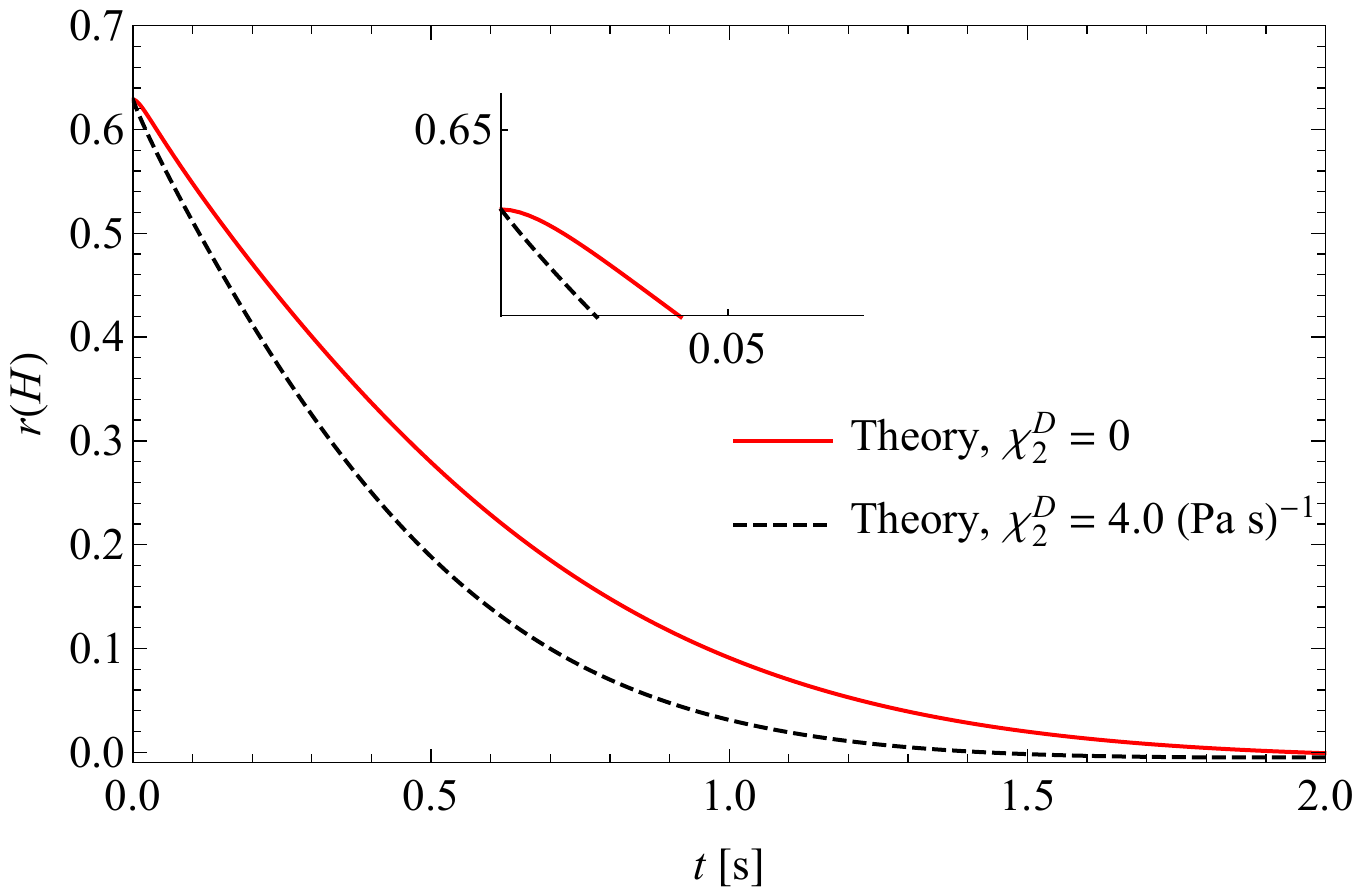}
	\caption{ (Color online) Normalized phase difference as a function of time at 5 mT, calculated with $\chi_2^D=0$ and $\chi_2^D=4.0$\,(Pa\,s)$^{-1}$.}
	\label{Bild23}
\end{figure}
Numerical calculations reveal that a strong dissipative cross-coupling causes the initial behavior of the normalized phase difference to be a linear function in time, Fig.~\ref{Bild23}, as found experimentally, Fig.~\ref{Bild25}.

\begin{figure}[htb]
	\includegraphics[width=3.3in]{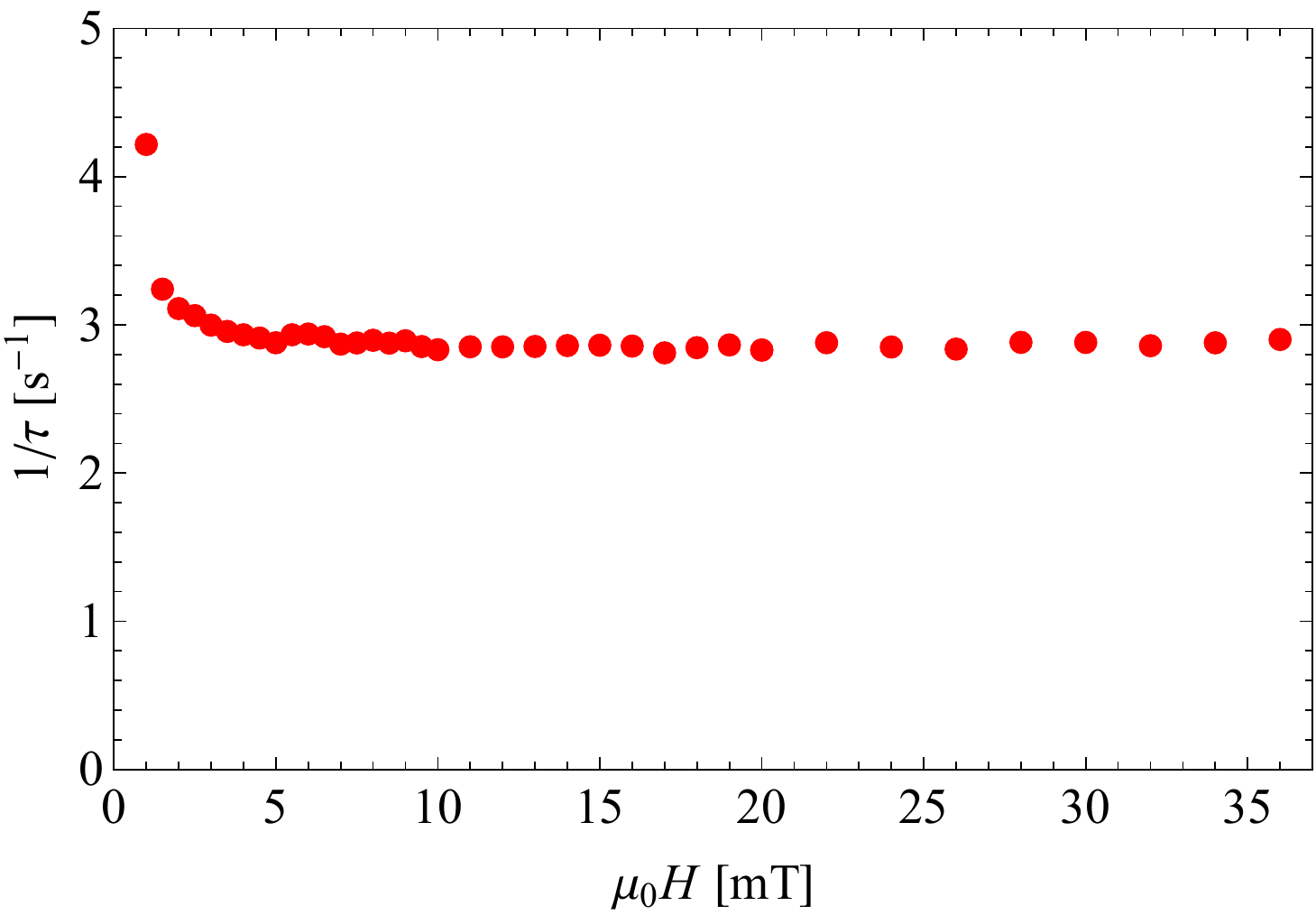}
	\caption{ (Color online) Experimental switch-off relaxation rate of the normalized phase difference as a function of the applied magnetic field.}
	\label{Bild26}
\end{figure}

\begin{figure}[htb]
\includegraphics[width=3.3in]{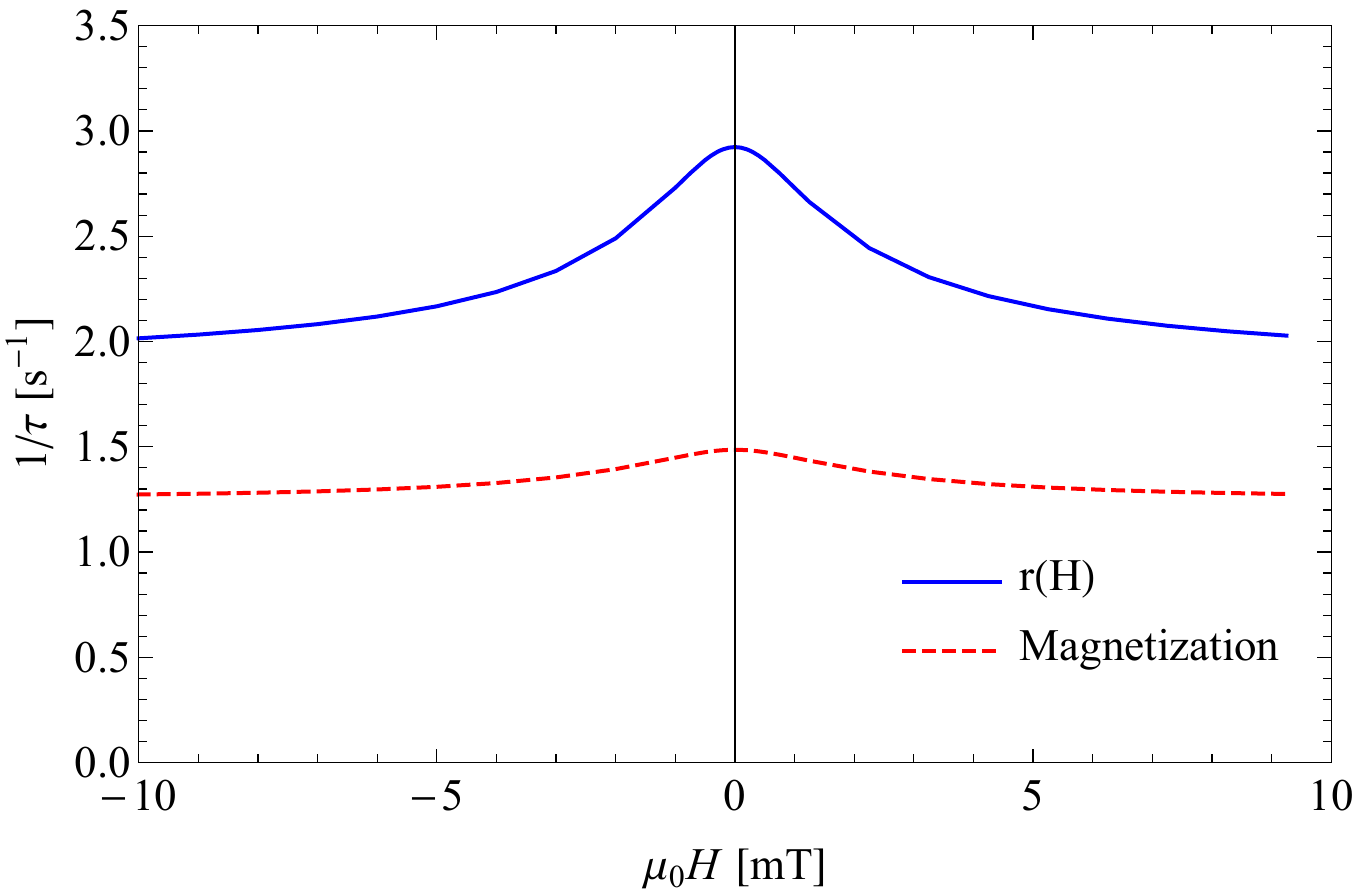}
\caption{ (Color online) Relaxation rate of the normalized phase difference and $z$ component of the magnetization after switching off the magnetic field of strength $\mu_0H$ at $\chi_2^D=4$\,(Pa\,s)$^{-1}$ and $\varphi_s=0$.}
\label{Bild28}
\end{figure}

To extract a relaxation time $\tau$ of the normalized phase difference we use an exponential function
\begin{equation}
	f(t)=f(0){\rm e}^{-t/\tau}.
\end{equation}
The relaxation rate $1/\tau$ for the experimental data is shown in Fig.~\ref{Bild26}.
It saturates at a finite value as one increases the magnetic field.
This is expected since the initial director and magnetization fields do not change much with magnetic field any more when the field is large.
In Fig.~\ref{Bild28} the relaxation rate of both the computed phase difference and the magnetization is shown. One can see that the relaxation rate of the magnetization is smaller than that of the normalized phase difference, owing to the fact that it is the director that is driven by the nonzero elastic force, while the magnetization only follows. This is true for all allowed values of the dynamic cross coupling parameters.

One can derive analytic formulas for the relaxation rate in the limit of low magnetic fields. With the assumption that the relaxation follows a simple exponential function, it is possible to extract the relaxation rate $1/\tau^{off}$ from the initial time derivative of the normalized phase difference,
\begin{equation}
	\label{eqfitx}
	r(H,t)\approx r(H,t=0)\left(1-\frac{t}{\tau^{off}}\right).
\end{equation}
Note that Eq.~(\ref{eqfitx}) is defined only when $r(H,t=0)\neq 0$.

One starts with the director quasi-current $\bf Y$, Eq.~(\ref{Y}). The response of the $z$ component of the director field is $n_z\approx  n_z(z,t=0)-Y_z(z,t=0)\,t$, which one uses in the equation Eq.~(\ref{eqfaza}) for the phase difference, 
\begin{equation}
	\label{tauoffdef}
	\frac{1}{\tau^{off}}=\frac{2k_0r_0(n_{e0}-n_o)}{\phi_0 r(H,t=0)}\int_{0}^d \mathrm{d}z\frac{n_z(z)Y_z(z) }{\left(1+\frac{n_{e0}^2-n_o^2}{n_o^2}n_z^2(z)\right)^{3/2}},
\end{equation}
where all $z$-dependent quantities are evaluated at $t=0$.
In the last step the integrand is expanded up to linear order in time and the relaxation rate in the low-magnetic-field limit is finally expressed as
\begin{equation}
	\label{tauoff}
	\frac{1}{\tau^{off}}=\frac{\left(1+r_0\varphi_s^2\right)\left[\mu_0HM_0\left(1+6\frac{\xi}{d}\right)+12\frac{K_1\varphi_s}{d^2}\right]\chi_2^D}{\frac{\mu_0HM_0d^2}{20K_1}\left(1+10\frac{\xi}{d}+30 \frac{\xi^2}{d^2}\right)+\left(1+6\frac{\xi}{d}\right)\varphi_s},
\end{equation}
which is linear in the dissipative cross-coupling coefficient $\chi_2^D$.

\begin{figure}[htb]
	\includegraphics[width=3.3in]{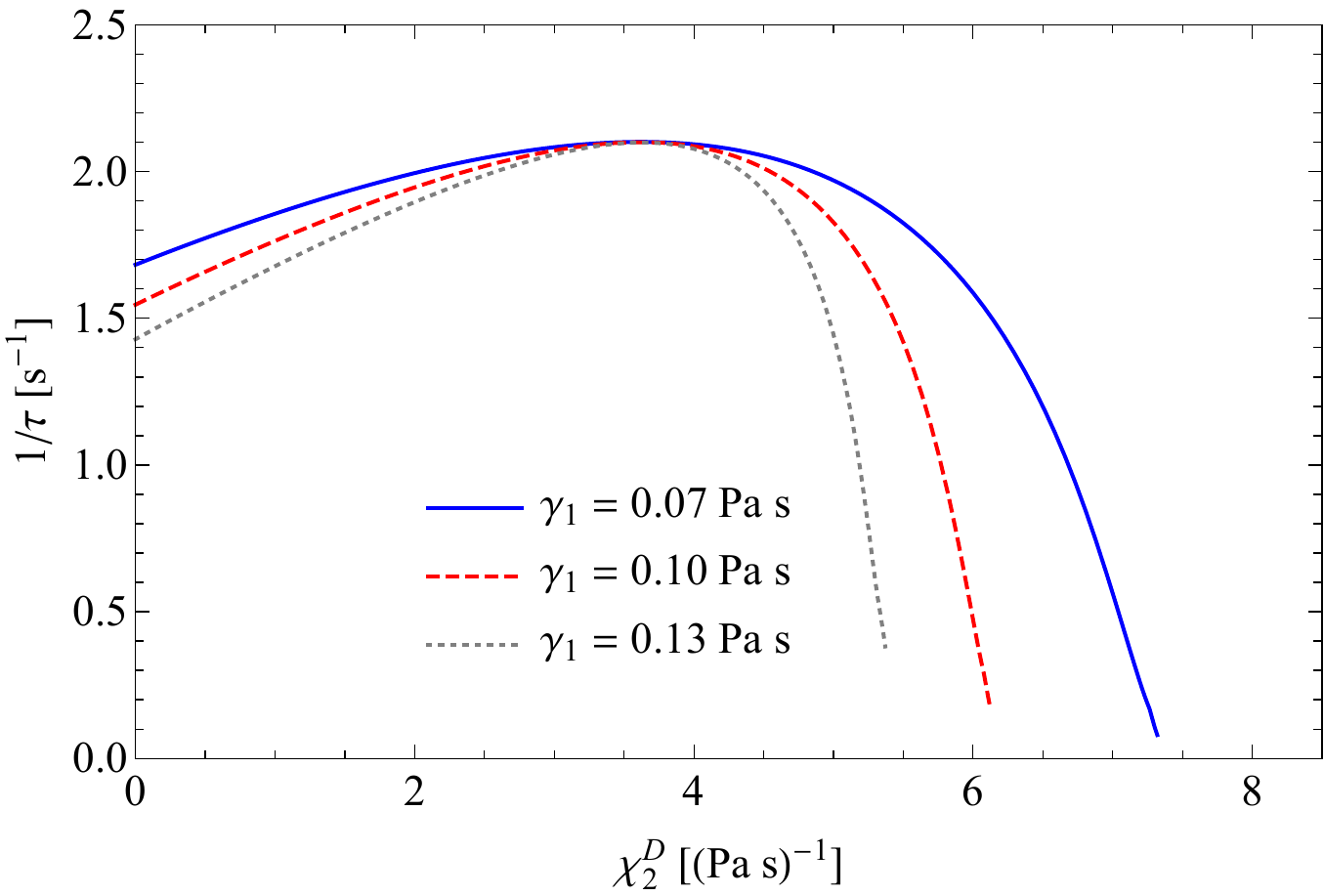}
	\caption{ (Color online) Relaxation rate at $\mu_0H=50$\,mT as a function of the dissipative coefficient $\chi_2^D$ for different values of the director rotational viscosity $\gamma_1$.}
	\label{Bild29}
\end{figure}

Not only does the dissipative cross-coupling make the switching process faster when switching on the field, this can be also true for switching off the field, Figs.~\ref{Bild23} and \ref{Bild29}.
%
Fig.~\ref{Bild29} shows the relaxation rate of the normalized 
phase difference at a high magnetic field as a function of the dissipative cross coupling coefficient $\chi_2^D$. 
As expected, the relaxation rate decreases with increasing rotational viscosity $\gamma_1$. 
The relaxation rate at first increases with increasing values of $\chi_{2}^D$, which seems also to be the case for small magnetic fields described by Eq.~(\ref{tauoff}). For values above approximately $\chi_2^D=3.5$\,(Pa\,s)$^{-1}$, the relaxation rate starts to decrease rather rapidly. This is in contrast with the field switch-on case, where the response is faster for increasing values of $\chi_2^D$. 

The increasing part of the dependence $\tau^{-1}(\chi_2^D)$ in Fig.~\ref{Bild29} is due to the director elastic forces, which drive the switch-off dynamics and also enter Eq.~(\ref{Mdot}) through the dissipative cross-coupling governed by $\chi_2^D$.
At higher values of $\chi_2^D$ one must however also consider the part of the thermodynamic forces corresponding to the static ($A_1$) coupling between the director and the magnetization. Focusing only on the director equation Eq.~(\ref{ndot}), one sees that the director relaxes towards the magnetization with a characteristic time set by the rotational viscosity and the static coupling ($A_1$). On the other hand, the positive value of $\chi_2^D$ has the opposite effect. While both fields eventually relax to the ground state parallel to $x$, the angle between them is decreasing slower and slower as the dynamic cross-coupling ($\chi_2^D$) gets larger. For small magnetic fields one can study the relaxation rate of the dynamic eigenmodes, Eq.~(\ref{eqflu}) of the next section. The value of $\chi_2^D$ above which the relaxation rate starts to decrease then reads
\begin{equation}
\chi_2^D=
  \begin{cases}
                                   \frac{A_1b_\perp^D}{A_1M_0^2+K(\pi/d)^2}& \text{if $\frac{1}{\gamma_1}>\frac{b_\perp^D}{M_0^2}$,} \\
                                  \frac{1}{\gamma_1} & \text{if $\frac{1}{\gamma_1}<\frac{b_\perp^D}{M_0^2}$.} 
  \end{cases}
\end{equation}
In our case $\frac{1}{\gamma_1}>\frac{b_\perp^D}{M_0^2}$ holds and the maximum is at $\chi_2^D \approx 3.5$\,(Pa\,s)$^{-1}$.

The switch-on case is different in that the dynamics is driven by the external magnetic field. If the external field is sufficiently high (large compared to $A_1M_0$), the static cross-coupling effects, which decrease the relaxation rate in the switch-off case through the increasing dynamic cross-coupling $\chi_2^D$, can be neglected and hence the relaxation rate is monotonically increasing with $\chi_2^D$.

%
%
%

\section{Fluctuations and light scattering}
\label{sec:flucts}

\noindent
Nematic liquid crystals appear turbid in sufficiently thick layers \cite{degennesbook}. The scattering of light is caused by strong director fluctuations which cause fluctuations in the dielectric tensor
\begin{equation}
\label{vareps}
	\varepsilon_{ij}=\varepsilon_\perp\delta_{ij}^\perp+\varepsilon_\parallel n_in_j,
\end{equation}
where $\varepsilon_\perp$ and $\varepsilon_\parallel$ are dielectric susceptibilities for the electric field perpendicular and parallel to the director, respectively. Fluctuations are easy to observe experimentally and are used to determine the viscoelastic properties of liquid crystals \cite{alenkaliqcryst}.


In this paper we derive the relaxation rates of the fluctuations without taking into account the effects of flow.  
Since the director is coupled to the magnetization, we now have two fluctuation modes for each director fluctuation mode of the usual nematic \cite{alenkanature}.

The fluctuating director and magnetization fields are linearized as
\begin{equation}
\begin{split}
\label{flukss}
\mathbf{n}&=\mathbf{n}_0+\delta \mathbf{n},\\
\mathbf{M}&=\mathbf{M}_0+M_0\delta \mathbf{m},
\end{split}
\end{equation}
where the equilibrium director $\mathbf{n}_0$ and magnetization $\mathbf{M}_0$ fields 
point in $x$ direction in which a magnetic field is applied, 
whereas fluctuations $\delta \mathbf{n}$ and $\delta \mathbf{m}$ are perpendicular, $\mathbf{n}_0\cdot\delta \mathbf{n}=\mathbf{M}_0\cdot\delta \mathbf{m}=0$.
The ansatz for the director fluctuations is 
\begin{equation}
\delta \mathbf{n}(\mathbf{r})=\frac{1}{V}\sum_{\mathbf{q}}\delta \mathbf{n}(\mathbf{q})e^{\mathrm{i}\mathbf{q}\cdot\mathbf{r}},
\end{equation}
where $\mathbf{q}=q_x \hat{\mathbf{e}}_x+q_y \hat{\mathbf{e}}_y+q_z \hat{\mathbf{e}}_z$ is the wave vector of the fluctuation. A similar ansatz is used for the fluctuations of the magnetization. In a confined system, the fluctuation spectrum generally depends on the interaction of the nematic with the surface \cite{alenkaliqcryst}.
For simplicity we will use the infinite anchoring limit, so that $q_z=n\pi/d, n\in \mathbb{N}$, while $q_x$ and $q_y$ are in principle arbitrary. For details regarding the anchoring effect we refer to Ref.~\cite{alenkaliqcryst}.


To understand the static light scattering experiments one must determine thermal averages of the fluctuations. This is done by finding linear combinations of the variables in terms of which the free energy functional Eq.~(\ref{f}) is expressed as a sum of quadratic terms, and making use of equipartition. Such linear combinations are uncorrelated (statistically independent).
A systematic way to perform this decomposition is to write the free energy of a fluctuation $\bf q$-mode as a quadratic form and find the corresponding eigenvalues and eigenvectors,
\begin{equation}
\label{eq1}
	F({\bf q})=\frac{1}{2}\mathbf{\delta x}({\bf q})^H \textsf{E}(\mathbf{q}) \mathbf{\delta x}({\bf q}),
\end{equation}
where 
$\mathbf{\delta x}({\bf q})=\{\delta n_z({\bf q}),\delta m_z({\bf q}),\delta n_y({\bf q}),\delta m_y({\bf q})\}$, in short $\mathbf{\delta x}({\bf q})\equiv\{n_z,m_z,n_y,m_y\}$, is the vector of the fluctuation amplitudes, $\textsf{E}({\bf q})$ is a self-adjoint matrix 
and superscript $^H$ is the conjugate transpose.

In lowest order of fluctuations, the contributions Eq.~(\ref{eq1}) of the free energy Eq.~(\ref{f}) are \cite{degennesbook}
\begin{align}
	F({\bf q})=\frac{1}{2V}\Bigg[&(K_1 q_y^2+K_2q_z^2+K_3q_x^2+A_1M_0^2)|n_y|^2\nonumber\\
	&+(K_1q_z^2+K_2 q_y^2+K_3q_x^2+A_1M_0^2)|n_z|^2\nonumber\\
	&+(K_1-K_2)q_zq_y(n_yn_z^*+n_y^*n_z)\nonumber\\
	&+(\mu_0HM_0+A_1M_0^2)(|m_y|^2+|m_z|^2)\nonumber\\
	&- A_1M_0^2(n_ym_y^*+n_y^*m_y+n_zm_z^*+n_z^*m_z)\Bigg].
\end{align}
For completeness (not needed here), the volume-integrated free energy is $F = \sum_{\bf q}F({\bf q})$.


Before giving the eigenvectors of the quadratic form $\textsf{E}$, we perform a rotation in the $yz$ plane, $(n_y,n_z)\to (n_1,n_2)$ and $(m_y,m_z)\to (m_1,m_2)$, where the new bases in this plane are $\{\hat{\bf e}^{n}_1,\hat{\bf e}^{n}_2\}$ and $\{\hat{\bf e}^{M}_1,\hat{\bf e}^{M}_2\}$. Vectors $\hat{\bf e}^{n}_2$ and $\hat{\bf e}^{M}_2$ are normal to ($\mathbf{q},\mathbf{n}_0$) and ($\mathbf{q},\mathbf{m}_0$) plane, respectively and vectors $\hat{\bf e}^{n}_1$ and $\hat{\bf e}^{M}_1$ are normal to $\hat{\bf e}^{n}_2$ and $\hat{\bf e}^{M}_2$, respectively. 
It should be emphasized that we are studying the case $\mathbf{n}_0||\mathbf{m}_0$, so the planes ($\mathbf{q},\mathbf{n}_0$) and ($\mathbf{q},\mathbf{m}_0$) are identical. In the confined system, this would not be the case if the external magnetic field were applied in any direction other than parallel to the initial homogeneous state.


A general fluctuation $\delta \mathbf{x}({\bf q})$ can be written as 
\begin{equation}
\label{amplit}
\delta \mathbf{x} = t_1 \mathbf{t}_1+p_1 \mathbf{p}_1+t_2 \mathbf{t}_2+p_2 \mathbf{p}_2,
\end{equation}
where $\mathbf{t}_1,\mathbf{t}_2,\mathbf{p}_1$, $\mathbf{p}_2$ are the eigenvectors of the quadratic form $\textsf{E}$ and $t_1,t_2,p_1$, $p_2$ are the amplitudes of these uncorrelated excitations. The eigenvectors are
\begin{eqnarray}
	\mathbf{t}_\alpha &=& a^t_\alpha \hat{\mathbf{e}}_\alpha^n + b^t_\alpha \hat{\mathbf{e}}_\alpha^M\label{eigenmod1} \\
	&=&\frac{Z_\alpha^-}{\sqrt{1+(Z_\alpha^-)^2}} \hat{\mathbf{e}}_\alpha^n-\frac{1}{\sqrt{1+(Z_\alpha^-)^2}}\hat{\mathbf{e}}_\alpha^M,\nonumber\\
	\mathbf{p}_\alpha &=& a^p_\alpha \hat{\mathbf{e}}_\alpha^n + b^p_\alpha \hat{\mathbf{e}}_\alpha^M \label{eigenmod2}\\
	&=&\frac{Z_\alpha^+}{\sqrt{1+(Z_\alpha^+)^2}} \hat{\mathbf{e}}_\alpha^n-\frac{1}{\sqrt{1+(Z_\alpha^+)^2}}\hat{\mathbf{e}}_\alpha^M,\nonumber
\end{eqnarray}
where
\begin{equation}
Z_\alpha^\pm=\frac{-\mu_0HM_0+K_\alpha q_ \perp^2+K_3q_x^2\pm s_\alpha}{2A_1M_0^2},
\label{Z_alpha}
\end{equation}
with $q_\perp^2=q_y^2+q_z^2$ and
\begin{equation}
	s_\alpha^2=4A_1^2M_0^4+\left(K_\alpha q_\perp^2+K_3q_x^2-\mu_0HM_0\right)^2.
\label{s_alpha}
\end{equation}

The excitation modes $\mathbf{t}_1$ and $\mathbf{p}_1$ are the analogues of the splay-bend mode in the usual NLCs, whereas $\mathbf{t}_2$ and $\mathbf{p}_2$ are the analogues of the twist-bend mode.
  
\begin{figure}[htb]
\includegraphics[width=3.3in]{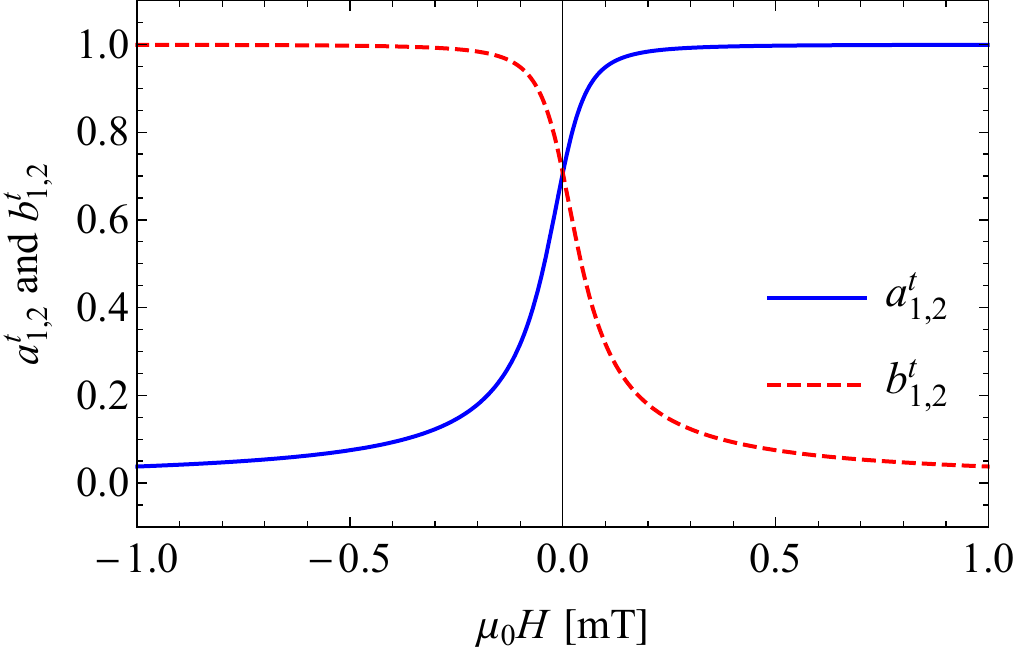}
\caption{ (Color online) The normalized coefficients Eq.~(\ref{eigenmod1}) of the eigenvectors $\mathbf{t}_1$ and $\mathbf{t}_2$ as a function of the applied magnetic field with $q_x=0$ and $q_\perp =\pi/2$; $K_1=K_2$. 
}
\label{Bild29a}
\end{figure}

\begin{figure}[htb]
\includegraphics[width=3.3in]{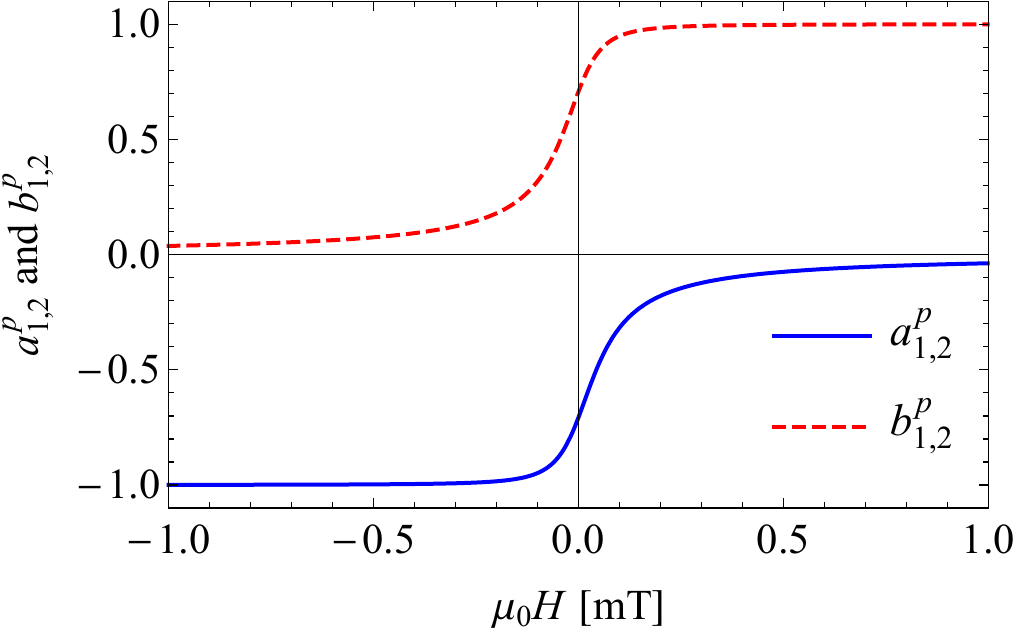}
\caption{ (Color online) The normalized coefficients Eq.~(\ref{eigenmod2}) of the eigenvectors $\mathbf{p}_1$ and $\mathbf{p}_2$ as a function of the applied magnetic field with $q_x=0$ and $q_\perp =\pi/2$; $K_1=K_2$. 
}
\label{Bild29b}
\end{figure}

It is found that in the limit of large magnetic fields these excitations become decoupled, i.e., one eigenvector only contains the fluctuation of the director field and the other the fluctuation of the magnetization field, Figs.~\ref{Bild29a} and \ref{Bild29b}.

The thermal averages of the squared amplitudes of the independent excitations read
\begin{align}
\label{thermal}
\langle |t_\alpha(\mathbf{q})|^2\rangle &= \frac{k_BTV}{\frac{1}{2}\left(2A_1M_0^2+\mu_0HM_0+K_\alpha q_\perp^2+K_3 q_x^2-s_\alpha\right)},\\
\langle |p_\alpha(\mathbf{q})|^2\rangle &= \frac{k_BTV}{\frac{1}{2}\left(2A_1M_0^2+\mu_0HM_0+K_\alpha q_\perp^2+K_3 q_x^2+s_\alpha\right)},\label{pms}
\end{align}
with $k_B$ the Boltzmann constant and $T$ the temperature, whereas their thermal cross-correlations are zero.

If $K_1=K_2$, the splay-bend $(\alpha=1)$ and the twist-bend $(\alpha=2)$ excitation modes have the same structure Eqs.~(\ref{Z_alpha})-(\ref{s_alpha}), Figs.~\ref{Bild29a} and \ref{Bild29b}, as well as the same energy and thermal amplitude Eqs.~(\ref{thermal})-(\ref{pms}).
The same is true in the degenerate case when $\mathbf{q}=q\,\hat{\mathbf{e}}_x$, i.e., for a pure bend excitation (in an unconfined system), where there is no difference between the modes $\alpha=1,2$ and the bases $\{\hat{\bf e}^{n}_1,\hat{\bf e}^{n}_2\}$ and $\{\hat{\bf e}^{M}_1,\hat{\bf e}^{M}_2\}$ are chosen arbitrarily in the $yz$ plane.


The space correlations 
are expressed as 
\begin{align}
\langle t_\alpha(\mathbf{r})t_{\alpha'}(\mathbf{r}')\rangle & = \frac{1}{V^2}\sum_{\mathbf{q},\mathbf{q}'}\langle t_\alpha(\mathbf{q})t_{\alpha'}(\mathbf{q}')\rangle e^{-\mathrm{i}(\mathbf{q}\cdot\mathbf{r}+\mathbf{q}'\cdot\mathbf{r}')}\nonumber \\
&=\frac{\delta_{\alpha,\alpha'}}{V^2}\sum_{\mathbf{q}}\langle t_\alpha(\mathbf{q})t_{\alpha}(-\mathbf{q})\rangle e^{-\mathrm{i}\mathbf{q}\cdot(\mathbf{r}-\mathbf{r}')},
\label{corr}
\end{align}
and similarly for $\langle p_\alpha(\mathbf{r})p_{\alpha'}(\mathbf{r}')\rangle$, whereas $\langle t_\alpha(\mathbf{r})p_{\alpha'}(\mathbf{r}')\rangle=0$. 
In the large magnetic field limit these correlations are
\begin{align}
\label{realcorel2}
\langle t_\alpha(\mathbf{r})t_\alpha(\mathbf{r}')\rangle &\approx \frac{k_BT}{4\pi K}\frac{1}{r}e^{-q_0r},\\
\langle p_\alpha(\mathbf{r})p_\alpha(\mathbf{r}')\rangle &\approx \frac{k_BT}{(2\pi)^3\mu_0HM_0} \delta(r),
\end{align}
where $r=|\mathbf{r}-\mathbf{r}'|$ and $q_0=\sqrt{A_1M_0^2/K}$.

In experiments one measures the intensity of the scattered light. To calculate this intensity we need an expression for the amplitude of the outgoing electric field. We start with an incident electric field $\mathbf{E}_i$, described by a plane wave: $\mathbf{E}=E_0\hat{\mathbf{i}}\,{\rm e}^{\mathrm{i}(\mathbf{k}_i\cdot \mathbf{r}-\omega t)}$, where $\mathbf{k}_i$ is the wave vector, $E_0$ the amplitude and $\omega$ the frequency of the incident light. We then proceed with a summation of the electric field contributions of the scattered light through the whole cell, treating every point $\bf r$ as a radiating dipole. Last, we project the electric field on the axis $\hat{\mathbf{f}}$ of the analyzer. The electric field amplitude of the scattered light is \cite{degennesbook}:
\begin{align}
\label{eqfield}
&E_f(\mathbf{q},t)\nonumber\\ &=\frac{E_0\omega^2}{c^2R}{\rm e}^{\mathrm{i}(\mathbf{k}_f\cdot\mathbf{r}'-\omega t)}\int_V\!\! \mathrm{d}^3r\,{\rm e}^{-\mathrm{i}\mathbf{q}\cdot \mathbf{r}}\hat{f}_i\, [\varepsilon_{ij}(\mathbf{r},t)-\delta_{ij}]\,\hat{i}_j \nonumber\\
&=\frac{E_0\omega^2}{c^2R}{\rm e}^{\mathrm{i}(\mathbf{k}_f\cdot\mathbf{r}'-\omega t)}\,\hat{f}_i \varepsilon_{ij}(\mathbf{q},t)\,\hat{i}_j,
\end{align}
where $\mathbf{k}_f$ is the wave vector of the scattered light, $R$ is the distance from the sample to the detector at ${\bf r}'$ and $\mathbf{q}=\mathbf{k}_f-\mathbf{k}_i$ is the fluctuation wave vector. In the last line of Eq.~(\ref{eqfield}) we discarded the Fourier contribution of $\delta_{ij}$, since it is nonzero only if $\mathbf{q}=0$. We have assumed that $R$ is large compared to the size of the scattering region which in turn is 
much larger than the wave length of the light, and that we are in the limit of small dielectric anisotropy.

In our calculations below, we will be using details of an experimental set-up usually used for measuring splay-bend fluctuations in a NLC, which in our geometry have $\delta{\bf n} = \delta n_z\hat{\bf e}_z$, $q_y = 0$, $\hat{\mathbf{e}}_2^{n,M}=\hat{\mathbf{e}}_y$ and $\hat{\mathbf{e}}_1^{n,M}=\hat{\mathbf{e}}_z$. In this case we have a polarizer and an analyzer that are both in the $xz$ plane. The polarizer $\hat{\mathbf{i}}$ is parallel to the $x$ axis, whereas the analyzer $\hat{\mathbf{f}}$ is at an angle $\zeta$ from the $x$ axis. 
In Eq.~(\ref{eqfield}), the projection of the fluctuating part of the dielectric tensor Eq.~(\ref{vareps}) reads
\begin{equation}
\hat{f}_i\, \varepsilon_{ij}(\mathbf{q},t)\,\hat{i}_j=\varepsilon_a f_z\delta n_z,
	\label{projection}
\end{equation}
where $f_z=\hat{\mathbf{f}}\cdot \hat{\mathbf{e}}_z$.
Using the expansion
\begin{equation}
\delta n_z=(t_1\mathbf{t}_1 +p_1\mathbf{p}_1) \cdot \hat{\mathbf{e}}_1^n,
\end{equation}
the scattering cross section $\sigma=\langle E_f^*(\mathbf{q},t)E_f(\mathbf{q},t)\rangle$ with ${\bf q}\cdot\hat{\bf e}_y=0$ is 
\begin{align}
\sigma&=\frac{\varepsilon_a^2\omega^4}{c^4}\langle |\delta n_z(\mathbf{q})|^2\rangle  f_z^2\nonumber \\&=\frac{\varepsilon_a^2\omega^4}{c^4}\left(C_1^+\langle |t_1(\mathbf{q})|^2\rangle+C_1^-\langle |p_1(\mathbf{q})|^2\rangle\right)  f_z^2,\label{scatt}
\end{align}
with the coefficient
\begin{align}
C_1^\pm&=\frac{(Z_1^\mp)^2}{1+(Z_1^\mp)^2}.
\end{align}
In the usual experimental set-up one observes two splay-bend modes, ${\bf t}_1$ and ${\bf p}_1$, as opposed to the usual NLC, where one observes only one splay-bend mode.

Asymptotic behaviors of the coefficients $C_1^+$ and $C_1^-$ at large magnetic fields, 
\begin{align}
	C_1^+&\asymp 1-\frac{A_1^2M_0^4}{(\mu_0HM_0)^2},\label{eqasc1}\\
	C_1^-&\asymp \frac{2(K_\alpha q_\perp^2+K_3 q_x^2)^2-3A_1^2M_0^4}{(\mu_0HM_0)^2},\label{eqasc2}
\end{align}
reveal that in the large magnetic field limit only the eigenmode $\mathbf{t}_1$ contributes to the scattering cross section Eq.~(\ref{scatt}).

The dynamics of the fluctuations is probed by dynamic light scattering, where one measures the time correlation of the light intensity $I(t)$,
\begin{equation}
g^{(2)}(t)=\frac{\langle I(0)I(t)\rangle}{\langle I(0) \rangle^2}.
	\label{g2}
\end{equation}
Assuming Gaussian fluctuations it follows
\begin{equation}
g^{(2)}(t)=1+\left|g^{(1)}(t)\right|^2,
	\label{g21}
\end{equation}
where
\begin{equation}
g^{(1)}(t)=\frac{\langle E_f^*(\mathbf{q},0)E_f(\mathbf{q},t)\rangle}{\langle |E_f(\mathbf{q},0)|^2 \rangle}\label{g1c}
\end{equation}
is the time correlation of the scattered light electric field.

To calculate the time dependence of the 
fluctuations, 
we first linearize the system of dynamic equations and determine the dynamic eigenmodes. Considering only the dissipative dynamics, Eqs.~(\ref{X})-(\ref{Y}), and using
$\delta \mathbf{n}=\delta n_1 \hat{\mathbf{e}}_1^n+ \delta n_2 \hat{\mathbf{e}}_2^n$, $\delta \mathbf{m}=\delta m_1 \hat{\mathbf{e}}_1^M+ \delta m_2 \hat{\mathbf{e}}_2^M$,
we find a 2$\times$2 homogeneous system for each $\alpha=1,2$,
\begin{equation}
\begin{split}
\label{eqflu}
\frac{1}{\tau}\delta n_\alpha&=\left[\frac{1}{\gamma_1}\left(K_\alpha q_\perp^2+K_3q_x^2+A_1M_0^2\right)-\chi_2^DA_1M_0^2\right]\delta n_\alpha \\ &+\left[A_1M_0^2\left(\chi_2^D-\frac{1}{\gamma_1}\right)+\chi_2^D\mu_0HM_0\right]\delta m_\alpha, \\
\frac{1}{\tau}\delta m_\alpha&=\left[-b_\perp^DA_1+\chi_2^D\left(K_\alpha q_\perp^2+K_3q_x^2+A_1M_0^2\right)\right]\delta n_\alpha\\ &+\left[b_\perp^DA_1 \left(1+\frac{\mu_0HM_0}{A_1M_0^2}\right)-\chi_2^DA_1M_0^2\right]\delta m_\alpha,
\end{split}
\end{equation}
which can be rewritten as
\begin{equation}
\label{flu2}
\left(\textsf{A}-\frac{1}{\tau}\textsf{I}\right)\begin{pmatrix}
\delta n_\alpha \\
\delta m_\alpha
 \end{pmatrix}=\mathbf{0}
\end{equation}
and has nontrivial solutions if $\mathrm{det}(\textsf{A}-\frac{1}{\tau}\textsf{I})=0$. 
%
The dynamic eigenmodes are the eigenvectors of the matrix $\textsf{A}$,
\begin{eqnarray}
\mathbf{t}_\alpha^h &=& c^{t}_\alpha \hat{\mathbf{e}}_\alpha^n + d^{t}_\alpha \hat{\mathbf{e}}_\alpha^M,\label{dyneigenmod1} \\
	\mathbf{p}_\alpha^h &=& c^{p}_\alpha \hat{\mathbf{e}}_\alpha^n + d^{p}_\alpha \hat{\mathbf{e}}_\alpha^M, \label{dyneigenmod2}    
\end{eqnarray}
where the components $c^{t}_\alpha, c^{p}_\alpha, d^{t}_\alpha, d^{p}_\alpha$ are functions of the static and dynamic material parameters and will not be given explicitly. 
It is important to realize that the dynamic fluctuation modes Eqs.~(\ref{dyneigenmod1})-(\ref{dyneigenmod2}) in general differ from the statistically independent excitation modes Eqs.~(\ref{eigenmod1})-(\ref{eigenmod2}).
If the reversible dynamics Eqs.~(\ref{XR})-(\ref{YR}) is included, a 4$\times$4 eigensystem is obtained coupling both $\alpha$'s. In that case splay-bend and twist-bend dynamic modes are no longer decoupled and each eigenmode spans all directions $\{\hat{\bf e}_1^ {n,M}, \hat{\bf e}_2^{n,M}\}$.

The time dependence of a fluctuation is first expressed in terms of the dynamic eigenmodes Eqs.~(\ref{dyneigenmod1})-(\ref{dyneigenmod2}), which are then further expressed by the uncorrelated excitation modes Eqs.~(\ref{eigenmod1})-(\ref{eigenmod2}).
Using Eqs.~(\ref{eqfield})-(\ref{projection}) and expressing $\delta n_z(t)$ of the splay-bend fluctuation as just explained, the electric field time correlation Eq.~(\ref{g1c}) becomes
\begin{equation}
|g^{(1)}(t)|=\frac{D_1^+(t)\langle |t_1(\mathbf{q},0)|^2\rangle +D_1^-(t)\langle |p_1(\mathbf{q},0)|^2\rangle }{C_1^+\langle |t_1(\mathbf{q},0)|^2\rangle+C_1^-\langle |p_1(\mathbf{q},0)|^2\rangle},
\end{equation}
where 
\begin{eqnarray}
D_1^+(t)&=& (\mathbf{t}_1 \cdot \hat{\mathbf{e}}_1^n)^2f_{\rm I}(t) +(\mathbf{t}_1 \cdot \hat{\mathbf{e}}_1^n)(\mathbf{t}_1 \cdot \hat{\mathbf{e}}_1^M)f_{\rm II}(t),\nonumber\\
D_1^-(t)&=& (\mathbf{p}_1 \cdot \hat{\mathbf{e}}_1^n)^2f_{\rm I}(t) +(\mathbf{p}_1\cdot \hat{\mathbf{e}}_1^n)(\mathbf{p}_1 \cdot \hat{\mathbf{e}}_1^M)f_{\rm II}(t).\nonumber\\
\end{eqnarray}
The functions $f_{\rm I}(t)$ and $f_{\rm II}(t)$ are expressed using the components $c^{t}_1, c^{p}_1, d^{t}_1, d^{p}_1$ and the relaxation times of the dynamic eigenmodes denoted by $\tau_1^t$ and $\tau_1^p$: 
\begin{align}
f_{\rm I}(t)&=\frac{c^{t}_1 d^{p}_1 {\rm e}^{-t/\tau_1^t}-c^{p}_1 d^{t}_1 {\rm e}^{-t/\tau_1^p}}{c^{t}_1 d^{p}_1-c^{p}_1 d^{t}_1},\\
f_{\rm II}(t)&=\frac{d^{t}_1d^{p}_1({\rm e}^{-t/\tau_1^t}- {\rm e}^{-t/\tau_1^p})}{c^{t}_1 d^{p}_1-c^{p}_1 d^{t}_1}.
\end{align}

In the limit of large magnetic fields one gets $D_1^\pm \to C_1^\pm {\rm e}^{-t/\tau_1^t}$. Taking into account also the large magnetic field dependence of the coefficients $C_1^\pm$, Eqs.~(\ref{eqasc1})-(\ref{eqasc2}), the intensity correlation function Eq.~(\ref{g21}) is a single exponential 
\begin{equation}
g^{(2)}(t)=1+{\rm e}^{-2t/\tau_1^t}.
\end{equation}

\begin{figure}[h]
\centering
\includegraphics[width=3.3in]{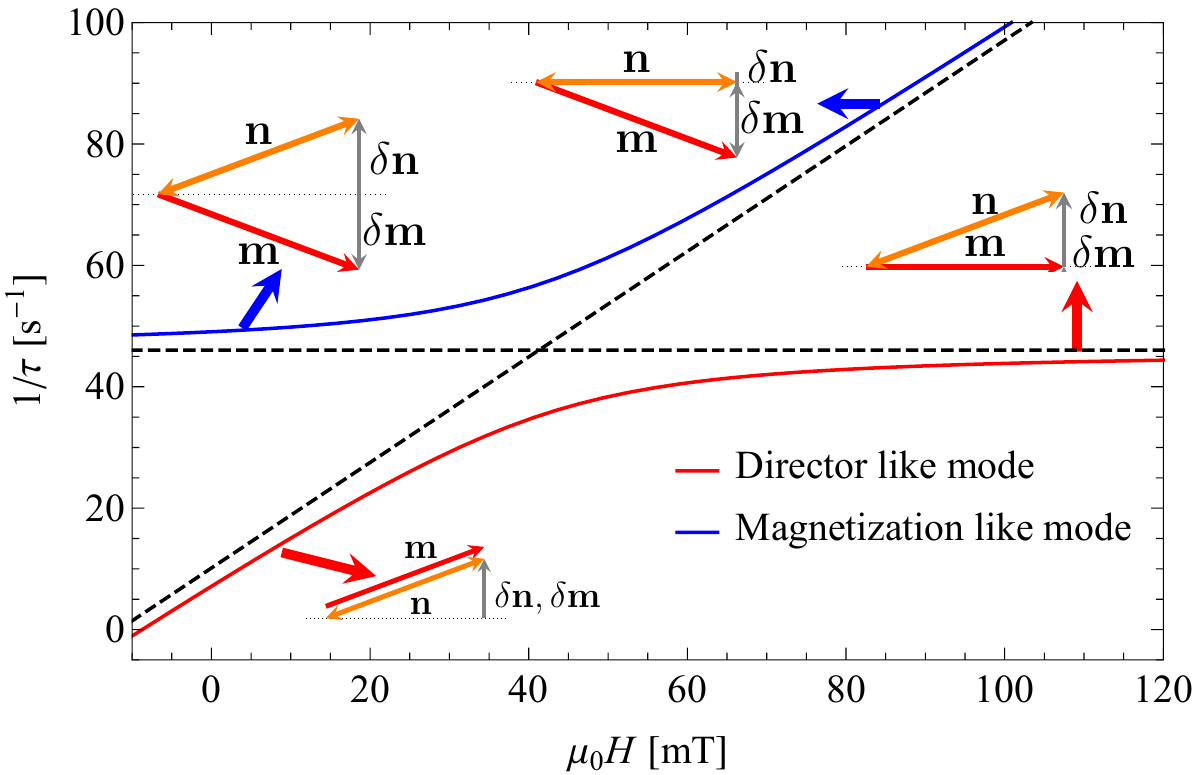}
\caption{Relaxation rates of almost pure bend fluctuations ($q_x\gg q_\perp$) and the corresponding dynamic eigenmodes as a function of the applied magnetic field. The dashed lines represent the limiting behavior of the relaxation rates, described by Eqs.~(\ref{lim1}) and (\ref{lim2}). For clarity, a smaller value of the rotational viscosity was used to make the asymptotic behavior set in sooner.}
\label{Bild30}
\end{figure}

It is found that the dynamics of the eigenmodes $\mathbf{t}_\alpha^h$ slows down ($\tau_\alpha^t \to \infty$) at a negative critical magnetic field, here given for ${\bf q}=q_z \hat{\bf e}_z$:
\begin{equation}
\label{bcrit}
\mu_0H_c^{(\alpha)}=-\frac{A_1M_0K_\alpha q_z^2}{K_\alpha q_z^2+A_1M_0^2}.
\end{equation}
Negative value of the critical magnetic field means that it is pointing in the direction opposite to the magnetization. If the applied magnetic field is more negative than the critical field, the magnetization 
starts to reverse. 
In NLCs, $K_2<K_1$ usually holds and it is the twist mode $\mathbf{t}_2^h$ that slows down at a less negative magnetic field. With the smallest wave number $q_z = \pi/d$ we get $\mu_0H_c^{(2)}=-2.5$\,mT.

In Fig.~\ref{Bild30} we present the magnetic field dependence of the relaxation rate of almost pure bend ($q_x\gg q_\perp$) fluctuations. 
We also depict the corresponding eigenmodes at a small positive field and at large magnetic fields. 

For a general fluctuation, 
in the limit of large magnetic fields the relaxation rate of the faster (magnetization-like) ${\bf p}_\alpha^h$ mode is proportional to the applied magnetic field (Fig.~\ref{Bild30} presents the bend fluctuation as an example),
\begin{eqnarray}
\label{lim1}
\frac{1}{\tau_\alpha^p}&=&\frac{A_1(b_\perp^D-\chi_2^DM_0)^2+(\chi_2^DM_0)^2(K_\alpha q_\perp^2+ K_3q_x^2)}{b_\perp^D}\nonumber \\ &+&\frac{b_\perp^D}{M_0}\mu_0H.
\end{eqnarray}
The relaxation rate of the slower (director-like) ${\bf t}_\alpha^h$ mode saturates at a finite value (Fig.~\ref{Bild30} presents the bend fluctuation as an example),
\begin{equation}
\label{lim2}
\frac{1}{\tau_\alpha^t}=\frac{A_1M_0^2+(K_\alpha q_\perp^2+K_3q_x^2)}{\gamma_1}\left(1-\frac{(\chi_2^DM_0)^2\gamma_1}{b_\perp^D}\right).
\end{equation}

It is also illuminating to study the relaxation rates of general fluctuations at zero magnetic field, $H=0$. Expanding the relaxation rates to second order in $q_x$ and $q_\perp$ one gets
\begin{eqnarray}
\frac{1}{\tau_\alpha^p}&=&\frac{A_1M_0^2}{\gamma_1}\left(1-2\chi_2^D\gamma_1+\frac{b_\perp^D\gamma_1}{M_0^2}\right)\nonumber\\
&+&\frac{(K_\alpha q_\perp^2+K_3q_x^2)\Xi_p}{\gamma_1},\label{zero1}\\
\frac{1}{\tau_\alpha^t}&=&\frac{(K_\alpha q_\perp^2+K_3q_x^2)\Xi_t}{\gamma_1},\label{zero2}
\end{eqnarray}
where
\begin{align}
\Xi_p&= \frac{(\chi_2^D\gamma_1-1)^2M_0^2}{b_\perp^D\gamma_1+(1-2\chi_2^D\gamma_1)M_0^2},\\
\Xi_t&=\frac{\gamma_1(b_\perp^D-(\chi_2^DM_0)^2\gamma_1)}{b_\perp^D\gamma_1+(1-2\chi_2^D\gamma_1)M_0^2}.
\end{align}
>From Eqs.~(\ref{zero1}) and (\ref{zero2}) one can see that the relaxation rate $1/\tau_\alpha^p$ of the faster (optic) mode ${\bf p}_\alpha^h$ stays finite in the limit $\mathbf{q}\to 0$. The slower mode ${\bf t}_\alpha^h$ is on the other hand acoustic, i.e., $1/\tau_\alpha^t \to 0$ as $\mathbf{q}\to 0$. 



\section{Summary and Perspective}
\label{sec:summary}

\noindent
In the present extensive study we have presented detailed 
experimental and theoretical investigations
of the dynamics of the magnetization and the director in a ferromagnetic 
liquid crystal in the absence of flow. 
We have shown that a dissipative cross-coupling between 
these two macroscopic variables, which has been determined quantitatively,
is essential to account for the experimental results also for the compound 
E7 as a nematic solvent for the ferromagnetic nematic phase. 
Before, this was demonstrated for 5CB as a nematic solvent \cite{tilenshort}.
We also find that all the experimental results presented here for E7 
complement well and are consistent with the previous ones using 5CB as the nematic component.
Remarkably, the dissipative cross-coupling ($\chi_2^D$) found for the E7-based ferromagnetic nematic liquid crystal is about a factor of 5 smaller than that of the 5CB-based, while the dissipative coefficient of the magnetization ($b_\perp^D$) is (only) twice as large.
This leads to an interesting suggestion for future 
experimental work, namely to address the question of which molecular features 
determine the strength of this dissipative cross-coupling.
The nematic phases of 5CB and E7, respectively, show one qualitatively different
feature: the nematic phase of 5CB is well-known to favour the formation of
transient pair-like aggregates \cite{cladis1981} because of its nitrile group,
while such tendencies are reduced in E7 since it is mixture of four different 
compounds and also
contains a terphenyl. A natural experiment to study 
these features in more detail would be to investigate the dependence   
of the dissipative cross-coupling on the magnetic particle concentration 
on one hand and to investigate mixtures
of the nematic solvents 5CB and E7 on the other to learn more about
the coupling mechanisms between the nematic order and the magnetic order. 

We have also analyzed the consequences of an out-of-plane dynamics, i.e., out of
the plane spanned by the magnetic field and the 
spontaneous magnetization. We give predictions for both, the azimuthal 
angles of director and magnetization as well as for 
the intensity change related to the reversible dynamic cross-coupling
terms between the two order parameters, the magnetization and the director. 
We find that from both measurements a value for the reversible cross-coupling
terms can be extracted. 
 
>From the present analysis the next steps in this field appear
to be quite well-defined.
First of all, the incorporation of flow effects appears to be highly desirable,
both from a theoretical as well as from an experimental point of view.
Early experimental results in this direction have been described in 
Ref.~\cite{alenkaapl}, where it has been shown that viscous effects can be tuned
by an external magnetic field of about $10^{-2}$\,T by more than a factor of two. 
>From a theoretical perspective questions like the analogues of the Miesowicz
viscosities and flow alignment are high on the priority list \cite{tilenflow}. 

Moreover, it will be important to realize, although perhaps 
experimentally challenging, 
a nematic or cholesteric liquid crystalline version of uniaxial
magnetic gels and rubbers \cite{collin2003,bohlius2004}. 
Cross-linking a ferromagnetic nematic
would give rise to the possibility to obtain a soft ferromagnetic gel,
opening the door to a new class of magnetic complex fluids.
This way one could combine the macroscopic degrees of freedom 
of the first liquid multiferroic, namely the ferromagnetic nematic
liquid crystal, with the strain field as well as with relative rotations.
In a step towards this goal, we will derive macroscopic dynamic
equations generalizing those for uniaxial magnetic gels and
ferronematics to obtain the macroscopic dynamics for ferromagnetic 
nematic and cholesteric gels \cite{tilengel}.

\vspace*{4mm}

\begin{acknowledgments}
Partial support of this work by H.R.B.,
H.P., T.P. and D.S. through the Schwerpunktprogramm SPP 1681 'Feldge\-steuerte
Partikel-Matrix-Wechselwirkungen: Erzeugung, skalen�bergreifende Modellierung
und Anwendung magnetischer Hybridmaterialien' of the Deutsche Forschungsgemeinschaft
is gratefully acknowledged, as well as the support of the Slovenian
Research Agency, Grants N1-0019, J1-7435 (D.S.), P1-0192 (A.M. and N.O.) and P2-0089 (D.L.). N.S. thanks the ``EU Horizon 2020 Framework Programme for Research and Innovation'' for its support through the Marie Curie Individual fellowship No. 701558 (MagNem). We thank the CENN Nanocenter for use of the LakeShore 7400 Series VSM vibrating-sample magnetometer.
\end{acknowledgments}

\end{document}